%% file: main.tex
\documentclass[fleqn,usenatbib,useAMS]{mnras}

\usepackage{graphicx}	
\usepackage{amsmath}	
\usepackage{amssymb}	
\usepackage{multicol}        
\usepackage{bm}		
\usepackage{pdflscape}	
\usepackage[normalem]{ulem}

\usepackage[T1]{fontenc}
\usepackage{ae,aecompl}

\usepackage{newtxtext,newtxmath}

\title[\texttt{matryoshka}]{\texttt{matryoshka}: Halo Model Emulator for the Galaxy Power Spectrum}

\author[J. Donald-McCann et al.]{
Jamie Donald-McCann,$^{1}$\thanks{E-mail: jamie.donald-mccann@port.ac.uk}
Florian Beutler,$^{2}$
Kazuya Koyama,$^{1}$
Minas Karamanis$^{2}$
\\
$^{1}$Institute of Cosmology \& Gravitation, University of Portsmouth, Dennis Sciama Building, Portsmouth, PO1 3FX, UK\\
$^{2}$Institute for Astronomy, University of Edinburgh, Royal Observatory, Blackford Hill, Edinburgh EH9 3HJ, UK
}

\date{Accepted XXX. Received YYY; in original form ZZZ}

\pubyear{2022}

\begin{document}
\label{firstpage}
\pagerange{\pageref{firstpage}--\pageref{lastpage}}
\maketitle

\begin{abstract}
\input{sections/abstract}
\end{abstract}





\section{Introduction}
\input{sections/Intro}

\section{Emulator Design}
\label{sec:design}
\input{sections/design}

\subsection{Base Model}
\label{subsec:analytic_emu}
\input{sections/analyticEmu}

\subsection{Nonlinear Boost Emulator}
\label{subsec:boost_emu}
\input{sections/boostEmu}

\subsection{Neural Networks as Emulators}
\label{subsec:NNs}
\input{sections/neuralnets}

\section{Emulator Accuracy}
\label{sec:accuracy}
\input{sections/accuracy}

\section{Mock Full Shape Analyses}
\label{sec:fullshape}
\input{sections/fullshape}

\section{Discussion}
\label{sec:disc}
\input{sections/disc}

\section{Conclusions}
\label{sec:conclusion}
\input{sections/conclusions}

\section*{Acknowledgement}
\input{sections/acknowledgement}

\section*{Data Availability}
\input{sections/data_av}

\bibliographystyle{mnras}
\bibliography{ref}

\appendix
\section{Monitoring Convergence of MCMC Chains}
\label{sec:convergence}
\input{sections/convergence}

\section{Accuracy Requirements}
\label{sec:accu_reqs}
\input{sections/requirements}

\bsp	
\label{lastpage}
\end{document}

%% file: sections/abstract.tex
We present \texttt{matryoshka}, a suite of neural network based emulators and accompanying Python package that have been developed with the goal of producing fast and accurate predictions of the nonlinear galaxy power spectrum. The suite of emulators consists of four linear component emulators, from which fast linear predictions of the power spectrum can be made, allowing all nonlinearities to be included in predictions from a nonlinear boost component emulator. The linear component emulators includes an emulator for the matter transfer function that produces predictions in $\sim 0.0004 \ \mathrm{s}$, with an error of $<0.08\%$ (at $1\sigma$ level) on scales $10^{-4} \ h \ \mathrm{Mpc}^{-1}<k<10^1 \ h \ \mathrm{Mpc}^{-1}$. In this paper we demonstrate \texttt{matryoshka} by training the nonlinear boost component emulator with analytic training data calculated with HALOFIT, that has been designed to replicate training data that would be generated using numerical simulations. Combining all the component emulator predictions we achieve an accuracy of $< 0.75\%$ (at $1\sigma$ level) when predicting the real space nonlinear galaxy power spectrum on scales $0.0025 \ h \ \mathrm{Mpc}^{-1}<k<1 \ h \ \mathrm{Mpc}^{-1}$. We use \texttt{matryoshka} to investigate the impact of the analysis setup on cosmological constraints by conducting several full shape analyses of the real space galaxy power spectrum. Specifically we investigate the impact of the minimum scale (or $k_\mathrm{max}$), finding an improvement of $\sim 1.8\times$ in the constraint on $\sigma_8$ by pushing $k_\mathrm{max}$ from $k_\mathrm{max}=0.25 \ h \ \mathrm{Mpc}^{-1}$ to $k_\mathrm{max}=0.85 \ h \ \mathrm{Mpc}^{-1}$, highlighting the potential gains when using clustering emulators such as \texttt{matryoshka} in cosmological analyses.

%% file: sections/Intro.tex
The next generation of galaxy surveys such as the Dark Energy Spectroscopic Instrument \citep[DESI,][]{levi_desi_2013,desi_collaboration_desi_2016} and Euclid \citep{laureijs_euclid_2011} will map the spatial distribution of galaxies with unprecedented accuracy. Studying the distribution of galaxies through their two-point clustering statistics has proven to be a valuable tool to understand the expansion history and matter content of the universe as well as the nature of dark energy \citep{cole_2df_2005,percival_measuring_2007,alam_clustering_2017,eboss_collaboration_completed_2021}. Traditionally these analyses have focused on linear or mildly-nonlinear scales, and put constraints on a cosmological model through scaling parameters such as $f\sigma_8$ (avoiding the need to recompute the full shape of the power spectrum for each cosmology considered). These choices are motivated by difficulties in accurately and efficiently modelling small nonlinear scales, however these small scales are where we have the largest statistical power from measurements of galaxy clustering. Producing accurate theoretical predictions on these small nonlinear scales will be essential to extract the highest amount of information from upcoming surveys. 

To accurately model galaxy clustering on small scales we rely on numerical simulations of a cosmological volume \citep{dolag_simulation_2008,kuhlen_numerical_2012,schneider_matter_2016,vogelsberger_cosmological_2019}. However these numerical simulations come at considerable computational cost, making it impractical using their direct outputs when fitting theoretical models to observed data. Complex perturbative models based on 1-loop perturbation theory can provide accurate predictions for galaxy clustering up to $k \sim 0.3 \ h \ \textrm{Mpc}^{-1}$ \citep{foreman_precision_2016,ivanov_cosmological_2020,philcox_combining_2020}, and thanks to the \texttt{FFTlog} \citep{hamilton_uncorrelated_2000}  these predictions can be made with a tractable computational cost \citep{simonovic_cosmological_2018}. In principle accurate predictions can be made on smaller scales by going to 2-loop order. The matter power spectrum can be accurately predicted with 2-loop models up to $k \sim 0.6 \ h \ \textrm{Mpc}^{-1}$ \citep{senatore_bias_2015,senatore_ir-resummed_2015}. The increased number of nuisance parameters that comes with going to 2-loop, and difficulty in constructing bias models that are accurate on small scales make this challenging in practice. 

\textit{Cosmic emulation} is a method that allows for simulation outputs to be used indirectly at a reasonable computation cost. Generally speaking an emulator is a sophisticated interpolation scheme with the ultimate goal of producing fast predictions that accurately reproduce the output of the model they are designed to replace. Emulators require \textit{training} on a given set of example outputs, what is meant by 'training' when discussing an emulator depends on the specifics of the interpolation scheme implemented in the emulator. A popular choice for the interpolation scheme when developing an emulator for galaxy clustering is Gaussian process (GP) regression. GPs are non-parametric and robust against overfitting, but are generally limited to producing scalar outputs. Using GPs to construct an emulator requires methodological choices that accommodate these scalar predictions. There has been great success combining GPs with a dimensionality reduction procedure such as principal component analysis (PCA) \citep{habib_cosmic_2007,heitmann_coyote_2009,kwan_cosmic_2015,lawrence_mira-titan_2017,giblin_road_2019,nishimichi_dark_2019}. When constructing an emulator in this way the hyper parameters of multiple Gaussian processes are optimised to predict the weights of the principal components of the training set. Successfully carrying out a PCA on training data that is often noisy and in a high-dimensional parameter space is non-trivial. Gaussian processes can be used in isolation when constructing an emulator \citep{zhai_aemulus_2019,bird_emulator_2019,pedersen_emulator_2021}. In this case individual Gaussian processes are trained to predict the value of a summary statistic of interest for different elements of a data vector (avoiding the need to employ PCA), i.e. different $r_p$ bins of the projected correlation function as in \citet{zhai_aemulus_2019}. Neural networks (NNs) have proven to be a viable alternative to GPs \citep{agarwal_pkann_2012, agarwal_pkann_2014, alsing_speculator_2020}. NNs are more susceptible to overfitting but are capable of producing vector outputs, and they also scale better than GPs with larger training sets. This capability of producing vector outputs is advantageous for multiple reasons. If a data vector is very large optimising and deploying a GP for each element of that data vector can be computationally expensive. In addition to this, producing predictions for individual elements of a data vector ignores any possible correlations between these elements, while producing vector outputs can preserve some of these correlations.

When developing emulators for the nonlinear matter power spectrum, it has proven useful to emulate a \textit{nonlinear boost} rather than emulating the matter power spectrum directly (see e.g.~\citealt{euclid_collaboration_euclid_2019,euclid_collaboration_euclid_2020, angulo_bacco_2020}).
This nonlinear boost is given by
\begin{equation}
    B(k) = \frac{P_{nl}(k)}{P_L(k)},
    \label{eq:boost}
\end{equation}
with $P_{nl}(k)$ being the nonlinear power spectrum that would be measured from a simulation, and $P_L(k)$ is a prediction of the power spectrum coming from linear theory. The major benefit of emulating this nonlinear boost rather than the power spectrum itself is that the boost has a simpler functional form than the power spectrum. We expect $B(k)\approx1$ on large scales where linear theory holds. The simpler functional form of the boost (and consequently lower dynamic range on large scales) leads to a higher prediction accuracy. A downside of this \textit{boosting} method is that the linear theory predictions can often create a bottleneck. The prediction of the boost coming from the emulator is generally orders of magnitude faster than the linear theory prediction. Several recent works have applied the idea of emulation to predictions coming from linear theory \citep[][]{arico_accelerating_2021,mootoovaloo_kernel-based_2021,mancini_itcosmopower_2021}. These linear theory emulators have uses beyond the boosting method mentioned above as they can also be used to speed up any cosmological analysis that requires a linear theory prediction for the power spectrum. In \citet{arico_accelerating_2021} for example, they emulate the linear matter power spectrum prediction to be used in a Lagrangian perturbation theory model for galaxy clustering.

For this work we aim to apply the boosting method mentioned above to the galaxy power spectrum. We also apply emulation to the linear theory predictions, allowing us to exploit the gain in prediction accuracy coming from the boosting method, whilst maintaining the fastest possible prediction speed. The paper is divided as follows; section \ref{sec:design} describes the suite of emulators that make up \texttt{matryoshka}. The tests we conduct to evaluate the prediction accuracy of each of the component emulators are outlined in section \ref{sec:accuracy}. Section \ref{sec:fullshape} describes several mock power spectrum full shape (FS) analyses that are designed to asses how the obtained level of prediction accuracy impacts parameter constraints when using \texttt{matryoshka} to do model fitting, and how the level of constraint is impacted by the minimum scale included in the analysis. Section \ref{sec:disc} contains short discussions about how the boost predicted by the emulator can be used to absorb complex small scale physics that is difficult to model analytically.
We conclude in section \ref{sec:conclusion}.

%% file: sections/design.tex
\texttt{matryoshka} is made up of a suite of emulators, predictions from each of these emulators are combined to make a prediction for the nonlinear galaxy power spectrum. The goal of \texttt{matryoshka} is to exploit the gain in accuracy that comes from emulating the nonlinear boost rather than the power spectrum directly, whilst maintaining the maximum possible prediction speed by also emulating several quantities that significantly increase the speed of the linear theory predictions for the galaxy power spectrum (hereafter the base model).

This section will describe the design of each of the emulators that make up \texttt{matryoshka}; the emulated components of the base model are outlined in section \ref{subsec:analytic_emu}, section \ref{subsubsec:analytic_data} describes the parameter space covered by the base model and how the training data for the base model can be focused on a particular suite of simulations, and the details of the galaxy halo connection model and nonlinear boost emulator are covered in section \ref{subsec:boost_emu}.

%% file: sections/analyticEmu.tex
For our base model we use the halo model (HM) framework \citep{cooray_halo_2002,murray_thehalomod_2020}. The HM is a very popular framework that has been used extensively when making predictions for galaxy clustering. We use the HM as it will allow us to very easily produce equivalent predictions from the base model and from a halo catalogue coming from a numerical simulation.

In the HM framework the clustering of galaxies is split into two regimes, describing the small and large scale clustering respectively

\begin{equation}
    P(k) = P_{1h}(k) + P_{2h}(k)\, .
    \label{eq:1h+2h}
\end{equation}
In the context of galaxy clustering these two terms are given by
\begin{equation}
    P_{gg,1h}(k) = \int n(M)\frac{\langle N(N-1)|M \rangle}{n^2_g}|u_g(k|M)|^2 dM,
    \label{eq:p_1h}
\end{equation}
and
\begin{equation}
    P_{gg,2h}(k) = P_L(k)\left[ \int n(M)b_h(M)\frac{\langle N|M \rangle}{n_g}u_g(k|M) dM \right]^2\, ,
    \label{eq:p_2h}
\end{equation}
with $n(M)$ being the halo mass function, $\langle N|M \rangle$ and $\langle N(N-1)|M \rangle$ being the expected number of galaxies and galaxy pairs respectively for a given halo mass $M$, $b_h(M)$ being the halo bias, $u_g(k,M)$ being the profile that satellite galaxies follow within a halo (in Fourier space), and $P_L(k)$ being the linear matter power spectrum. Throughout this work we assume that the satellite galaxies directly follow the \citet*[][here after NFW]{navarro_structure_1996} profile of their host halo. The NFW profile has the form
\begin{equation}
    \rho(r|M)\propto\frac{1}{r\frac{c(M)}{r_h(M)}\left[1+r\frac{c(M)}{r_h(M)}\right]^2}\ ,
\end{equation}
where $r_h(M)$ is the radius of a halo with mass $M$ assuming spherical halos, and $c(M)$ is the halo concentration-mass relation. With the use of fitting functions for $n(M)$, $b_h(M)$, and $c(M)$ \citep{duffy_dark_2008}, both terms in equation \ref{eq:1h+2h} can be calculated analytically at a relatively small, but non-negligible computational cost. The nonlinear boost component emulator (discussed in section \ref{subsec:boost_emu}) produces predictions in $\sim 0.0004 \ \mathrm{s}$\footnote{This prediction time and all others referred to in this paper are based on predictions made on a laptop with a 2.7 GHz Quad-Core Intel Core i7 CPU.}. For comparison this is $\sim 200 \times$ faster than the linear power spectrum calculation needed for equation \ref{eq:p_2h}. If we were not to emulate the base model the prediction time of the nonlinear galaxy power spectrum would be dominated by the prediction time of the base model. Hence we are motivated to emulate several components of equations \ref{eq:p_1h} and \ref{eq:p_2h}.

\subsubsection{Base Model Component Emulation}
We emulate four quantities that allow us to greatly increase the speed of the base model predictions. Those are the matter transfer function $T(k)$, the mass variance $\sigma(M)$, the logarithmic derivative of the mass variance $\frac{d\text{ln}\sigma(M)}{d\text{ln}M} \equiv \mathcal{S}(M)$, and the linear growth function $D(z)$. To achieve the best accuracy for a prediction of the power spectrum from the HM, $T(k)$ is normally calculated using a Boltzmann code such as \texttt{CAMB} \citep{lewis_efficient_2000} or \texttt{CLASS} \citep{lesgourgues_cosmic_2011}. There are cheaper analytic alternatives, such as \citet{eisenstein_baryonic_1998}, but the accuracy on large scales from these cheaper alternatives is not high enough. When developing a boosting emulator it is important that there is good agreement on large scales between the base model prediction and the prediction coming from the numerical simulation, as this is what gives the small dynamic range for $B(k)$ on large scales. The transfer function enters equations \ref{eq:p_1h} and \ref{eq:p_2h} in multiple terms, such as the matter power spectrum
\begin{equation}
    P_L(k) = A_s k^{n_s} T^2(k)\, , 
    \label{eq:P_L}
\end{equation}
where $A_s$ is the amplitude of the primordial power spectrum, and $n_s$ is the spectral index. $T(k)$ also enters indirectly in the halo mass function $n(M)$ and halo bias $b_h(M)$ via the mass variance $\sigma(M)$, given by
\begin{equation}
    \sigma^2(M) = \frac{1}{2 \pi^2}\int_0^\infty k^2P_L(k)W^2(kM)dk\ ,
    \label{eq:sigma}
\end{equation}
and the logarithmic derivative $\mathcal{S}(M)$, given by
\begin{equation}
    \mathcal{S}(M) = \frac{3}{2 \pi^2 r^4 \sigma^2(M)}\int_0^\infty \frac{dW^2(kM)}{dM}\frac{P_L(k)}{k^2}dk\ .
    \label{eq:dnls}
\end{equation}
Although we could just emulate $T(k)$ and evaluate \ref{eq:sigma} and \ref{eq:dnls} directly using the $T(k)$ emulator predictions, these calculations add non-negligible time to the base model prediction, so we decide to emulate $\sigma(M)$ and $\mathcal{S}(M)$ in addition to $T(k)$. In the equations above $W(kM)$ is the Fourier transform of the window function, throughout this work we use a top-hat window function of the form
\begin{equation}
    W(kM) = 3\frac{\sin(kM)-kM\cos(kM)}{(kM)^3}\ .
\end{equation}
For our base model we use a \citet{tinker_toward_2008} halo mass function, with the form
\begin{equation}
    n(M) = \frac{\rho_0}{M^2}f_n[\sigma(M)]\left|\mathcal{S}(M)\right|\ ,
    \label{eq:dndm}
\end{equation}
where the function $f_n[\sigma(M)]$ is given by,
\begin{equation}
    f_n[\sigma(M)]=A\left[\left(\frac{b}{\sigma(M))}\right)^a+1 \right]\exp{\left(-\frac{c}{\sigma(M)^2}\right)}\ ,
    \label{eq:hmf_func}
\end{equation}
and the coefficients of the function above are calibrated against simulations in \citet{tinker_toward_2008}. For the base model we use a \citet{tinker_large_2010} halo bias with the form
\begin{equation}
    b_h(M) = 1-A\frac{\nu^a}{\nu^a+\delta_c^a}+B\nu^b+C\nu^c\ ,
    \label{eq:halo_bias}
\end{equation}
where $\nu=\delta_c/\sigma(M)$, $\delta_c=1.686$, and as with equation \ref{eq:hmf_func}, the coefficients of \ref{eq:halo_bias} are calibrated against simulations in \citet{tinker_large_2010}.

To avoid including the redshift $z$ as an extra input parameter for the linear component emulators we include all redshift dependence through the growth factor $D(z)$; $P(k,z) \propto D^2(z)$, $\sigma^2(M,z) \propto D^2(z)$, and $\mathcal{S}(M) \propto D^2(z)/D^2(z)$. This means we can emulate $T(k)$, $\sigma(M)$, and $\mathcal{S}(M)$ at redshift zero, and include all redshift dependence with an emulator for $D(z)$. These four component emulators are what make up the \textit{emulated} base model.

\subsubsection{Parameter Space \& Training Data}
\label{subsubsec:analytic_data}

\begin{figure}
	\includegraphics[width=\columnwidth]{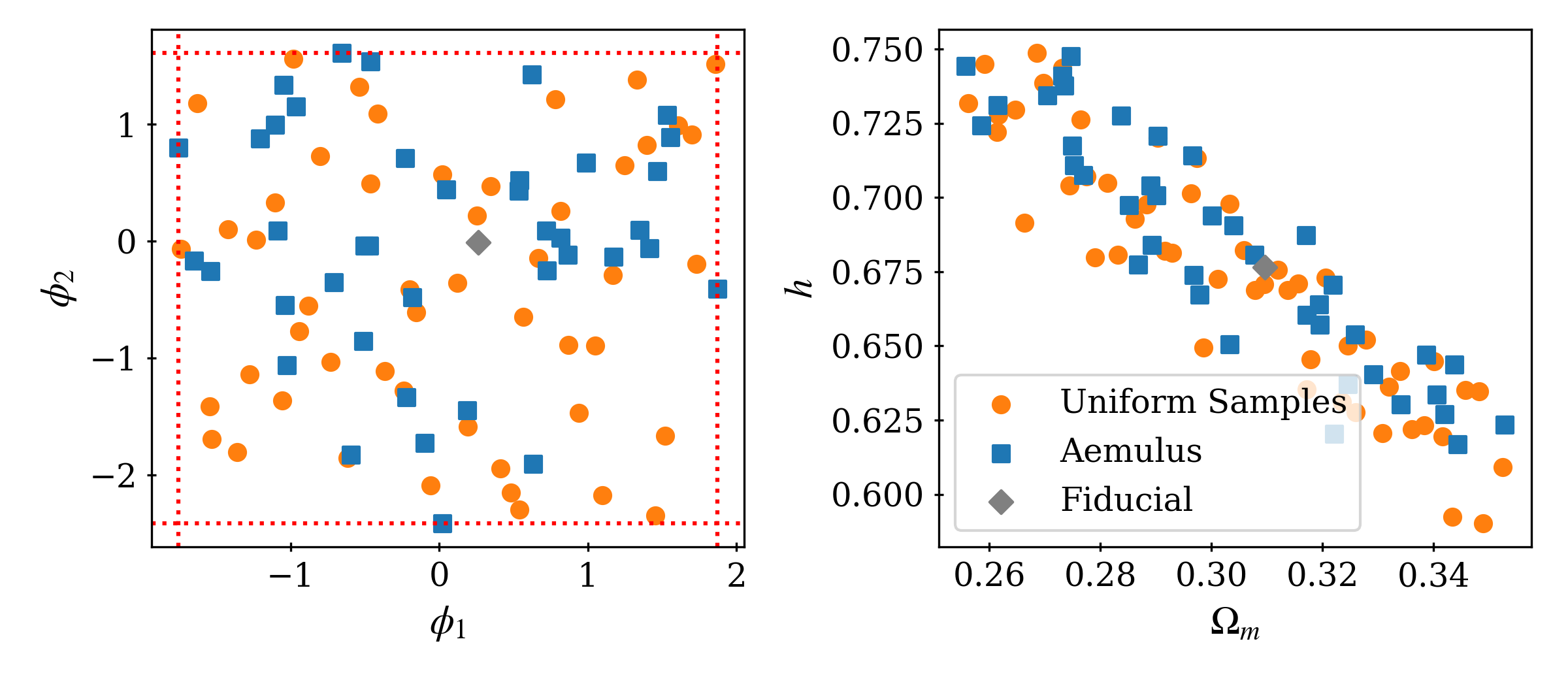}
    \caption{Visualisation of the procedure of focusing the \texttt{matryoshka} training data on a suite of simulations. The left panel shows the 40 Aemulus training cosmologies (blue squares) and our fiducial cosmology (grey diamond) in the uncorrelated latent space. The orange circles show 50 random samples in the latent space, and the red dotted lines show the boundaries in the latent space defined by the extreme values of the Aemulus suite. The right panel shows how these 50 samples are distributed in the cosmological parameter space.}
    \label{fig:latent_space}
\end{figure}

\begin{figure*}
	\includegraphics[width=\linewidth]{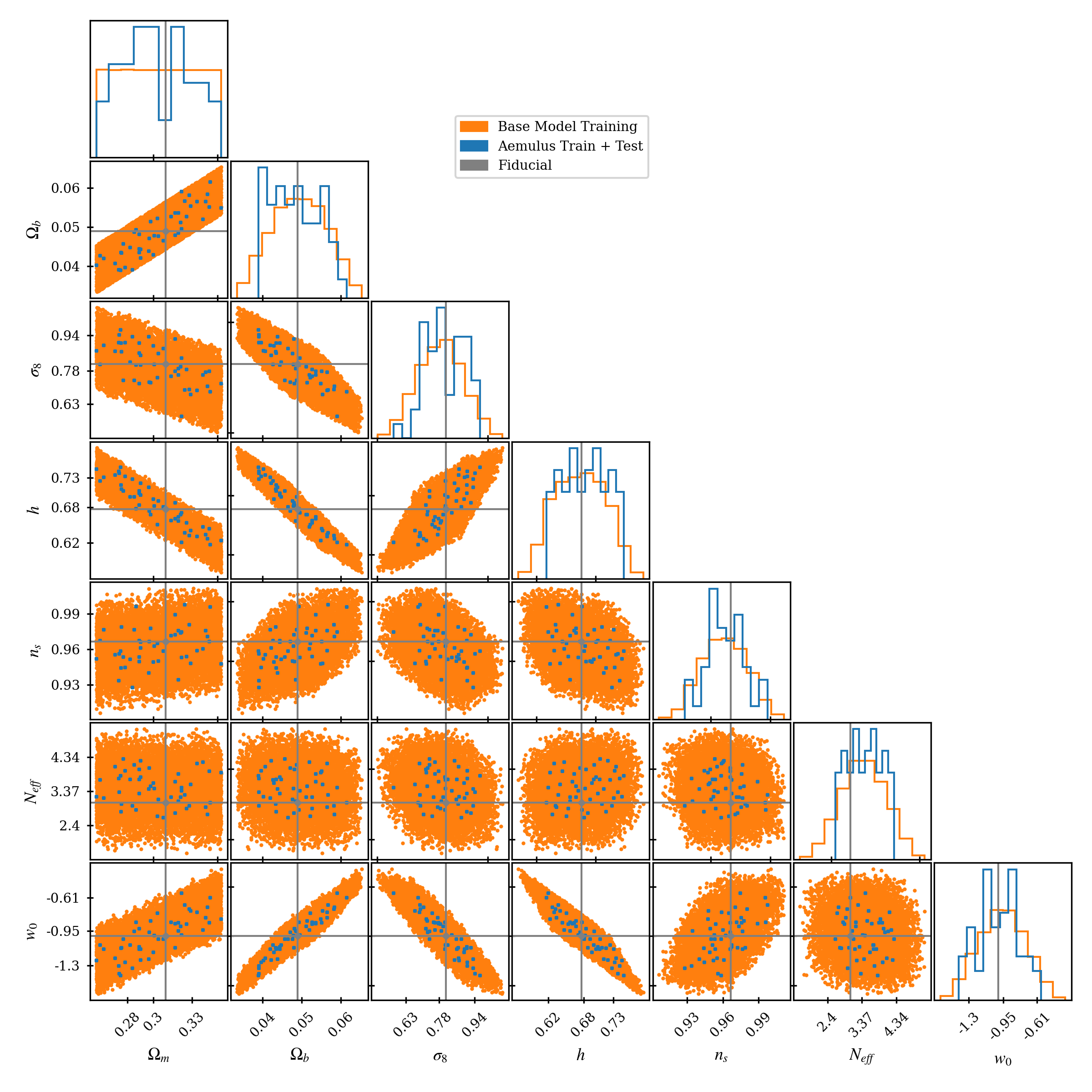}
    \caption{1D and 2D projections of the training space for the cosmological parameters of our model. The blue points and histograms show the Aemulus training and test cosmologies. The sampling of these training cosmologies is influenced by results from CMB, BAO, and SN experiments \citep[see section 2][]{derose_aemulus_2019}. The orange points and histograms show the training data for the base model of \texttt{matryoshka}. This training data has been generated to cover the same region of the parameter space as the Aemulus training cosmologies by uniformly sampling from an uncorrelated latent space defined by the Aemulus training cosmologies (see section \ref{subsubsec:analytic_data}). The grey point and solid lines show the location of the cosmology used to generate the mock power spectrum for the full shape analyses described in section \ref{sec:fullshape}.}
    \label{fig:linear_emu_space}
\end{figure*}

For this work we focus on the context where the base model will be trained to be used alongside a suite of numerical simulations, such as the Abacus Cosmos \citep{garrison_abacus_2018}, Aemulus \citep{derose_aemulus_2019}, Dark Quest \citep{nishimichi_dark_2019}, or Quijote \citep{villaescusa-navarro_quijote_2020} simulations. We choose to consider the Aemulus simulations. This publicly available simulation suite totals 75 dark matter only simulations, each with a volume of $( 1.05 \ \mathrm{Gpc} \ h^{-1})^3$. 40 of these simulations form a training set that has already been used to successfully train an emulator for the correlation function in redshift space \citep{zhai_aemulus_2019}. This emulator has recently been used to measure the growth rate from the eBOSS LRG sample \citep{chapman_completed_2021}. These 40 simulations sample a seven dimensional cosmological parameter space, with parameters $\{\Omega_m, \Omega_b, \sigma_8, h,  n_s, N_{\text{eff}}, w_0 \}$. The samples are selected to uniformly cover a $4\sigma$ region on these seven parameters coming from analysis of cosmic microwave background (CMB), baryon acoustic oscillations (BAO), and supernovae measurements \citep[for details see section 2 of][]{derose_aemulus_2019}. To focus our training data for our base model on the Aemulus simulations we calculate the covaraince of these 40 training simulations, we then use the Cholesky decomposition $\Sigma = \mathbfit{L}\mathbfit{L}^\mathrm{H}$ to decompose the covaraince matrix into the lower triangle and corresponding conjugate transpose (indicated by the superscript H). This lower triangle is used to transform the samples that make up the Aemulus training set into an uncorrelated latent space, shown in figure \ref{fig:latent_space}. We generate 10,000 samples uniformly in each dimension of this latent space, with intervals defined by the latent space samples corresponding to the Aemulus simulations. The lower triangle \textbf{\textit{L}} can then be used to transform these latent space samples back into the cosmological parameter space. Sampling the parameter space in this way allows us to focus only on the region where we will have simulated training data to combine with our base model, and not extend the training space (at the determent of prediction accuracy) to regions that will not be covered by simulations. It should be noted that in the case where the simulations uniformly cover the parameter space, such as the Dark Quest simulations, this procedure is not necessary. All of the 1D and 2D projections of the cosmological parameter space are shown in figure \ref{fig:linear_emu_space}. We can see that the 10,000 samples cover the region sampled by the Aemulus simulations whilst minimising sampling in regions where there are no simulations. Figure \ref{fig:linear_emu_space} also shows that our fiducial cosmology (based on the most recent Planck $\Lambda$CDM TT, TE, EE + lowE + lensing + BAO analysis, see section \ref{subsec:mock_cmass}) that will be used in the various analysis tests in section \ref{sec:fullshape} is located roughly at the centre of this seven dimensional parameter space.

The 10,000 samples are split into training, validation, and test subsamples (with 6400, 1400, 2000 samples respectively). The components of the emulated base model are trained using the training subsample, the validation subsample is used for model selection, and the prediction accuracy of the emulated base model is tested using the test subsample. Transfer functions and growth factors are calculated for the 10,000 samples using \texttt{CLASS} as implemented in the Python package \texttt{nbodykit} \citep{Hand_2018}, for 400 logarithmically spaced $k$-bins in the interval $10^{-4} <k<10 \ h \ \textrm{Mpc}^{-1}$ and at 200 linearly spaced redshifts in the interval $0<z<2$. These transfer functions are then used to calculate $\sigma^2(M)$ and $\mathcal{S}(M)$ with the Python package \texttt{hmf} \citep*{murray_hmfcalc_2013}.

\begin{table}
 \centering
 \begin{tabular}{c c c} 
 \hline
 Parameter & Range & Fiducial Value \\ 
 \hline
 $h$ & $[0.570, 0.780]$ & 0.6766 \\ 
 $\Omega_m$ & $[0.256, 0.353]$ & 0.30966 \\
 $\Omega_b$ & $[0.0334, 0.0653]$ & 0.04897 \\
 $w_0$ & $[-1.58, -0.322]$ & -1. \\
 $N_\text{eff}$ & $[1.61, 5.14]$ & 3.046 \\
 $\sigma_8$ & $[0.502, 1.07]$ & 0.8102 \\
 $n_s$ & $[0.906, 1.01]$ & 0.9665 \\
 \hline\hline
 Emulator & Parameters & Architecture\\
 \hline
 $T(k)$ & $\{\Omega_m, \Omega_b, h, N_\mathrm{eff}, w_0 \}$ & 5:300:300:300 \\
 $D(z)$ & $\{\Omega_m, \Omega_b, h, N_\mathrm{eff}, w_0 \}$ & 5:200:200:200 \\
 $\sigma(M)$ & $\{\Omega_m, \Omega_b, h, N_\mathrm{eff}, w_0, n_s, \sigma_8 \}$ & 7:200:200:500 \\
 $\mathcal{S}(M)$ & $\{\Omega_m, \Omega_b, h, N_\mathrm{eff}, w_0, n_s, \sigma_8 \}$ & 7:200:200:500 \\
\hline
\end{tabular}
\caption{Table defining the ranges for each of the parameters considered when constructing the base model emulators, along with the parameters used to calculate the mock observations for the fiducial full shape analysis in section \ref{sec:fullshape}. We also indicate which parameters are used by which emulator(s) and the architecture of each base model component emulator.}
\label{tab:linear_param}
\end{table}

%% file: sections/boostEmu.tex
\subsubsection{Halo Occupation Distribution Model}
A key ingredient when calculating the galaxy power spectrum in the HM framework is the halo occupation distribution (HOD). A HOD model is a probabilistic model that describes the probability of a dark matter halo of mass $M$ hosting $N$ galaxies of a given type. For this work we use the popular five parameter \citet{zheng_halo_2009} model. This model is split into two terms describing the occupation of central and satellite galaxies separately, where the expected central occupation is modelled as a smoothed step-function

\begin{equation}
    \langle N_\mathrm{cen}|M \rangle = \frac{1}{2}\mathrm{erfc}\left[\frac{\ln{M_\mathrm{cut}}-{\ln{M}}}{\sqrt{2}\sigma}\right]\, , 
\end{equation}
and the expected number of satellite galaxies is modelled as a power law
\begin{equation}
    \langle N_\mathrm{sat}|M \rangle =
    \begin{cases}
      0 & \text{if}\ M < \kappa M_\mathrm{cut}\, , \\
      \left(\frac{M-\kappa M_\mathrm{cut}}{M_1}\right)^\alpha & \text{if}\ M > \kappa M_\mathrm{cut}\, .
    \end{cases}
\end{equation}
$M_\mathrm{cut}$ defines the minimum mass for a halo to host a central galaxy, $\sigma$ defines to what extent the central step-function is smoothed, the product $\kappa M_\mathrm{cut}$ defines the minimum mass for a halo to host a satellite, $M_1$ defines the typical mass for a halo to host a satellite, and $\alpha$ defines how the expected number of galaxies increases with mass. We also impose the condition that a halo cannot host a satellite galaxy without first hosting a central galaxy, such that the expected total occupation is given by 
\begin{equation}
    \langle N|M \rangle = \langle N_\mathrm{cen}|M \rangle(1 + \langle N_\mathrm{sat}|M \rangle). 
\end{equation}

In order to calculate $P_{gg,1h}(k)$ (equation \ref{eq:p_1h}) we need to compute the expected number of pairs for a given halo mass $\langle N(N-1)|M \rangle$. As shown in section 3.1 of \citet{zheng_theoretical_2005} $\langle N(N-1)|M \rangle$ can be written as 
\begin{equation}
    \langle N(N-1)|M \rangle = 2\langle N_\mathrm{cen} N_\mathrm{sat}|M \rangle+\langle N_\mathrm{sat}(N_\mathrm{sat}-1)|M \rangle.
    \label{eq:cs_pairs}
\end{equation}
Under the assumption that the number of satellite galaxies follows a Poisson distribution we can write that $\langle N_\mathrm{sat}(N_\mathrm{sat}-1)|M \rangle=\langle N_\mathrm{sat}|M \rangle^2$, and following \citet{miyatake_cosmological_2020} we approximate $\langle N_\mathrm{cen} N_\mathrm{sat}|M \rangle=\langle N_\mathrm{cen}|M \rangle \langle N_\mathrm{sat}|M \rangle$ under the central condition.
This is a simple HOD model, and for this work this is the HOD model used in the emulated base model, in addition to being used to generate nonlinear power spectra that will train the boost emulator. It should be noted that there is no requirement for the galaxy halo connection model to be identical for emulated base model and the nonlinear boost component emulator. The only requirement is that it is possible to relate the two models such that power spectra predictions produced by the base model and those coming from the simulation agree on large scales (see section \ref{subsec:boost_absorb}).

In section \ref{sec:fullshape} we conduct a series of mock power spectrum FS analyses for a BOSS CMASS \citep{dawson_baryon_2012} style power spectrum, as such we define the extent of the HOD parameter space to be the same as that from \citet{kwan_cosmic_2015}. This parameter space was designed to cover the HOD models from the analyses of BOSS CMASS galaxies by \citet{white_clustering_2011}. The ranges of the five HOD parameters are given in table \ref{tab:HOD_param}.

\begin{table}
 \centering
 \begin{tabular}{c c c} 
 \hline
 Parameter & Range & Fiducial Value \\ 
 \hline
 $\log{M_\mathrm{cut}}$ & $[12.9, 13.78]$ & 13.04 \\ 
 $\log{M_1}$ & $[13.5, 14.7]$ & 14.05 \\
 $\sigma$ & $[0.5, 1.2]$ & 0.94 \\
 $\kappa$ & $[0.5, 1.5]$ & 0.93 \\
 $\alpha$ & $[0.5, 1.5]$ & 0.97 \\
 \hline
 \end{tabular}
\caption{Table defining the ranges for each of the HOD parameters used to train the nonlinear boost emulator, along with the parameters used to calculate the mock observations for the fiducial full shape analysis in section \ref{sec:fullshape}. The extent of the HOD parameter space matches that of \citet{kwan_cosmic_2015} and is designed to cover the results of \citet{white_clustering_2011}. These results from \citet{white_clustering_2011} also define our fiducial HOD parameters.}
\label{tab:HOD_param}
\end{table}

\subsubsection{Analytic Training Spectra}
\label{subsubsec:fake_sim}

For this work we aim to introduce \texttt{matryoshka}, and demonstrate how the \texttt{matryoshka} emulated base model can be focused on a suite of simulations to be used along side the nonlinear boost component emulator that has been trained on data coming from that same suite of simulations. Generating this training data from simulations is cheaper than running the simulations themselves, but still comes at considerable computational cost. With this in mind we opt to generate the training data for the nonlinear boost component emulator with HALOFIT \citep{takahashi_revising_2012}. Nonlinear training data is calculated via equation \ref{eq:1h+2h}, with $P_L$ in equation \ref{eq:p_2h} being replaced with the nonlinear matter power spectrum, with nonlinearities introduced via HALOFIT. This allows us to very quickly generate data for the nonlinear boost component emulator, and demonstrate \texttt{matryoshka}. We take steps to replicate the scenario where simulated training data is used, such as introducing noise into the HALOFIT training data that would be present in simulated training data, and only generating training data for the 40 Aemulus training cosmologies. In future work we will train this nonlinear boost component emulator directly on simulated training data. 

Following a similar procedure to \citet{zhai_aemulus_2019} we sample the HOD parameter space 50 times for each cosmology, resulting in 2000 training samples for the nonlinear boost component emulator. When calculating power spectra associated with these cosmological and HOD parameters we only include scales that would be accessible from a simulation box. The smallest $k$-mode is defined by the fundamental mode of the simulation box 
\begin{equation}
    k_\textrm{fund} = \frac{2\pi}{L_\textrm{box}},
\end{equation}
with $L_\textrm{box}$ being the length of one side of the simulation box. For a simulation from the Aemulus suite with $L_\textrm{box}=1.05 \ \textrm{Gpc} \ h^{-1}$ we have $k_\textrm{fund} \approx 6 \times 10^{-3} \  h \  \mathrm{Mpc}^{-1}$. To determine the highest $k$-mode that we want to emulate we consider the smallest scales that can be accurately represented by a dark matter only (DMO) N-body simulation. Many works have studied the impact of baryonic effects on the dark matter power spectrum, by comparing the dark matter power spectrum measured from a DMO simulation and hydrodynamic counterpart \citep{schneider_quantifying_2019,arico_modelling_2020,debackere_impact_2020}, and although there is some disagreement between different hydrodynamical codes and baryonic feedback models as to how strong the impact is on the dark matter power spectrum, there is general agreement that for scales where $k \gtrsim 1 \  h \  \mathrm{Mpc}^{-1}$ the impact on the dark matter power spectrum is >1\%. With this in mind we consider $k\approx 1 \  h \  \mathrm{Mpc}^{-1}$ the smallest possible scale that can be accurately modelled without including baryonic effects in the simulation and thus the smallest scale we want to emulate. With these limitations that would come from simulated training data in mind we limit the analytic training data coming from HALOFIT to 127 $k$-values from $0.012\  h \  \mathrm{Mpc}^{-1}\lesssim k \lesssim 1.152\  h \  \mathrm{Mpc}^{-1}$ \footnote{These values correspond to bin centres from a hypothetical power spectrum measurement covering $k_\mathrm{fund.}<k<k_\mathrm{Nyq.}/2$, where $k_\mathrm{Nyq.}=N_\mathrm{mesh} \pi / L_\mathrm{box}$ and $N_\mathrm{mesh}=1024$ such that the smallest emulated scale covers at least $k_\mathrm{Nyq.}/2$.}.   

As mentioned above simulated training data comes with noise. One source of noise is the sample variance of the simulations in the suite. There are methods to reduce the impact of this sample variance on the training data. In the case of the Aemulus simulations the initial conditions for each of the training simulations are generated with different random seeds, which prevents the emulator learning the noise coming from a specific set of initial conditions. Running phase-matched simulations effectively removed this sample variance \citep{angulo_cosmological_2016,chuang_unit_2019,klypin_suppressing_2020}, however this method doubles the computational cost of producing the training simulations without increasing the density of the sampling in the training space. It is not clear if a suite of phase-matched simulations would outperform a suite with random phases but with twice the sampling density. Another source of noise in simulated training data comes from the procedure used to populate the simulation with galaxies. Populating simulated dark matter halos according to an HOD is a random process, as such multiple realisations of the same HOD will result in variation of the measured power spectrum. To try and remove some of this HOD realisation noise, multiple realisations are normally generated and the power spectrum from each of these realisation is averaged. This averaging will remove some but not all of the HOD realisation noise. To approximate this left over noise we take the average of 10 random draws from a multivariate Gaussian with mean being the smooth nonlinear power spectrum calculated with HALOFIT and covariance give by
\begin{equation}
    C_{ii} = \frac{(2\pi)^3}{V}\left[\frac{2(P(k_i) + n_g^{-1})^2}{4\pi k_i^2 dk}\right].
    \label{eq:gauss_cov}
\end{equation}

%% file: sections/neuralnets.tex
A \textit{neural network} (NN) is a machine learning algorithm structured into \textit{layers} and \textit{nodes}. The nodes take numeric values corresponding to the weighted sum of all the nodes in the previous layer. In the case where we have two layers, with $i$ and $j$ nodes respectively, each node of the second layer will take the value given by
\begin{equation}
    y_j = \sum_i W_{i,j}\ x_i\, .
    \label{eq:linear_NN}
\end{equation}
In the equation above $W_{i,j}$ are the weights connecting the nodes of the first layer $x_i$ to the $j$th node of the second layer.
From equation \ref{eq:linear_NN} it is clear that a NN structured in this way will only be able to learn functions that correspond to a linear combination of the inputs. To enable the NN to learn nonlinear functions of the inputs \textit{activation functions} are applied to each node
\begin{equation}
    y_j = f\left(\sum_i W_{i,j}\ x_i\right)\, ,
    \label{eq:nonlinear_NN}
\end{equation}
where $f(...)$ is the nonlinear activation function.
The nonlinearities introduced by these activation functions allow NNs to become \textit{universal approximators} and make them a good choice when constructing an emulator. When we train a NN given a set of input-output pairs, we adjust the weights $\vec{W} = \{ W_0, ..., W_n\}$ such that the output of the NN most closely matches the target function output. 

\begin{figure*}
	\includegraphics[width=\linewidth]{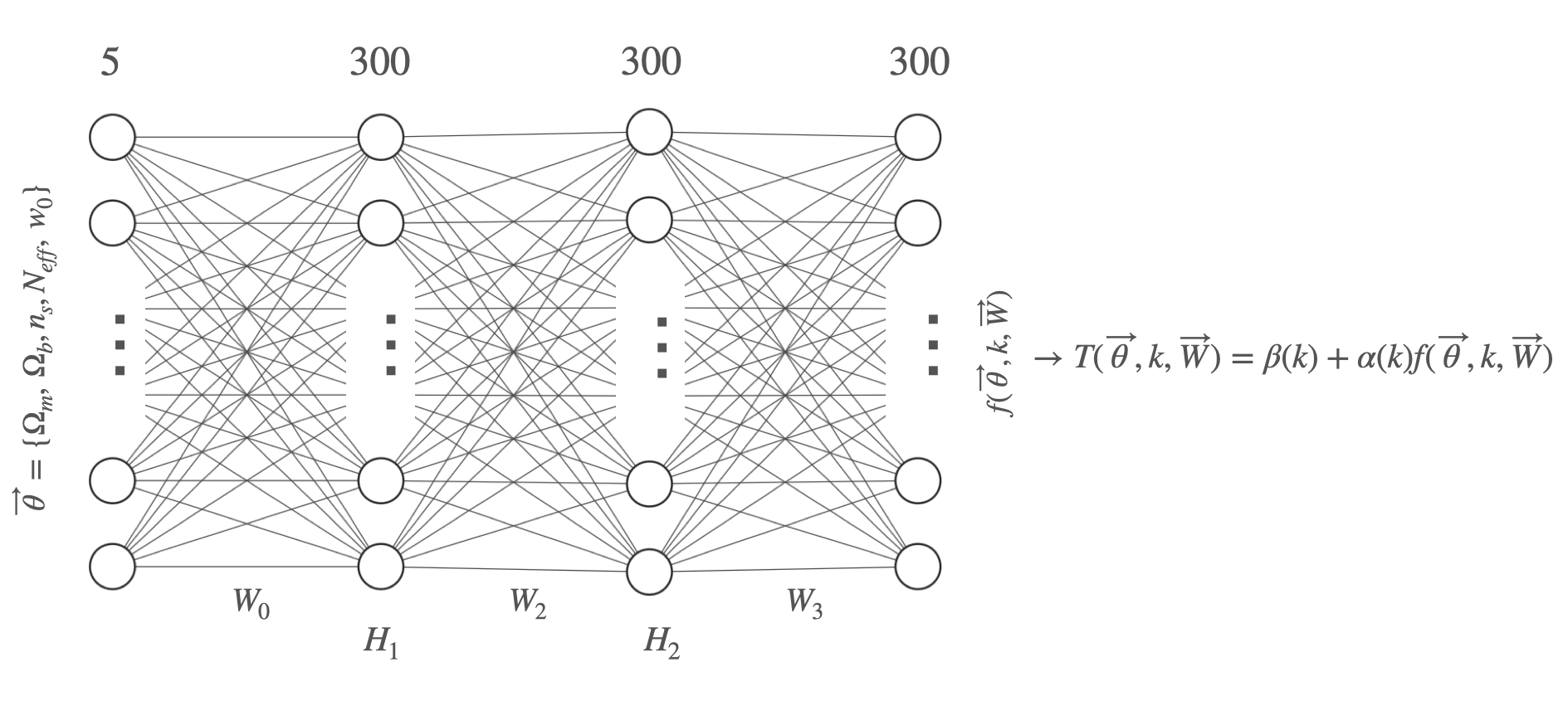}
    \caption{Schematic of the $T(k)$ component of the base model. The size of the input and output layers depend on the component being emulated, otherwise all components in the base model have a similar architecture (all base model component architectures are given in table \ref{tab:linear_param}). The nodes in both hidden layers $H_1$ and $H_2$ have the ReLU activation function applied, while there is no activation applied to the input or output layers. The functions $\alpha(k)$ and $\beta(k)$ post process the NN output for it to be interpreted as $T(k)$.}
    \label{fig:Temu_arch}
\end{figure*}

The number of input parameters, layers, nodes in each layer, and outputs define the \textit{architecture} of the NN. For this work we use NNs with a relatively simple architecture. A schematic visualising the architecture of the transfer function component of the emulated base model described in section \ref{subsec:analytic_emu} is shown in figure \ref{fig:Temu_arch}. We can summarise this architecture in a compressed way by writing 5:300:300:300, this allows us to immediately see that the NN has five input parameters, two hidden layers with 300 nodes each, and an output of 300 nodes. All components of the emulated base model share a very similar architecture to the transfer function, which are given in table \ref{tab:linear_param}. The nonlinear boost component emulator has an architecture of 12:200:200:127. The hidden layers in all the components of \texttt{matryoshka} have the ReLU activation applied to their nodes, whilst the input and output layers have no activation applied. These are not unique architectures. They were hand-tuned to achieve sub-percent accuracy from the \texttt{matryoshka} predictions on all scales considered (for a discussion on accuracy requirements see appendix \ref{sec:accu_reqs}). Whilst aiming for sub-percent accuracy we also try to maintain the simple architectures with a small number of layers and nodes. Simple NNs are less susceptible to overfitting and are more computationally efficient when producing predictions. Starting with a minimal architecture, a single hidden layer with 50 nodes, the numbers of nodes and/or layers is increased until the increase has minimal or negative impact on the optimised value for the loss function. Grid searches or even more sophisticated methods of architecture selection could result in higher prediction accuracy, however they were not necessary to achieve the desired level of accuracy for this work.

We apply preprocessing to the training data for each of the component emulators. The preprocessors all have the form
\begin{equation}
    f(x, \theta_i) = \frac{g(x, \theta_i)-\beta(x)}{\alpha(x)}\, ,
\end{equation}
where $f(x, \theta_i)$ is the processed target function for the $i$th sample in the training set, $g(x, \theta_i)$ is the unprocessed target function, and $\alpha(x)$ and $\beta(x)$ are functions that summarise the training data. With $\alpha(x_i) = \max\{ g(x_i, \theta) \} - \beta(x_i)$ and $\beta(x_i) = \min \{ g(x_i, \theta) \}$. The result of this preprocessing is that all the training data now lie in the interval $[0,1]$ for all target functions and at all scales. This has the effect that all scales contribute equally to the loss function, meaning that no scale is preferentially recovered as a consequence of the magnitude of the target function at that scale. The preprocessing also improves stability of the training procedure. Having all scales lying in the interval $[0,1]$ ensures that the weights don't get too large during training.

\begin{figure}
	\includegraphics[width=\columnwidth]{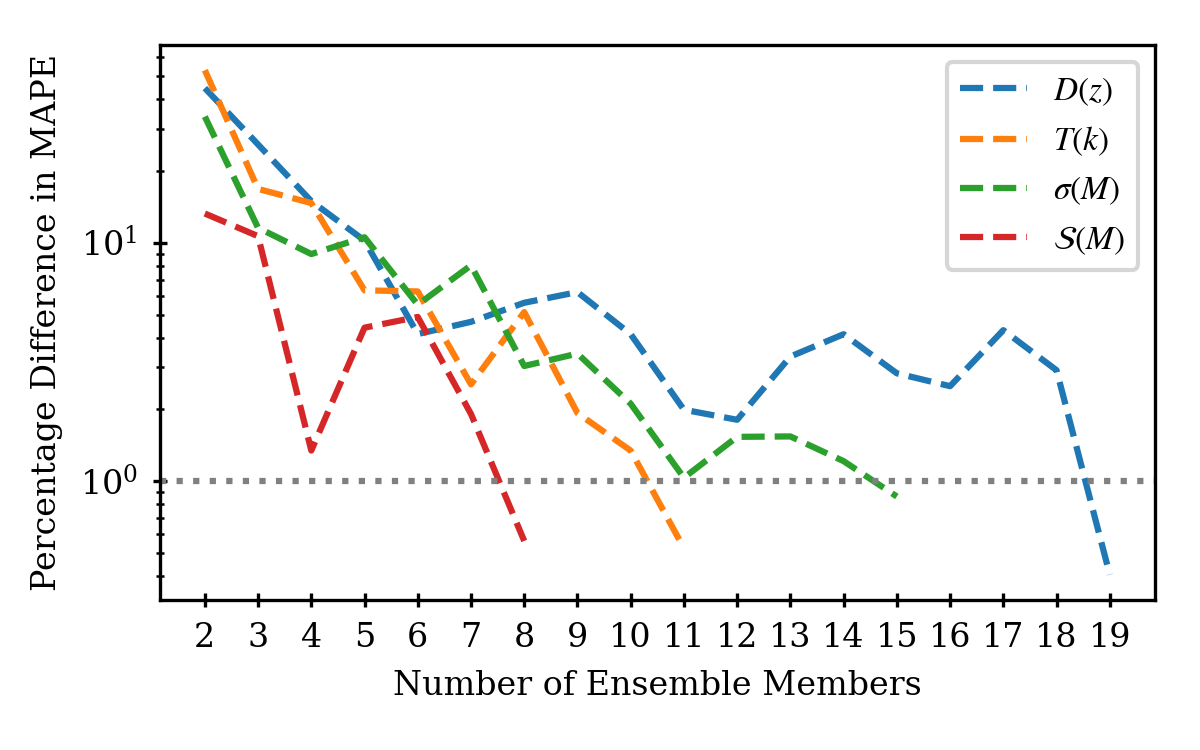}
    \caption{Plot showing the percentage difference in the mean absolute percentage error (MAPE) with increasing number of ensemble members when making predictions on the validation set for the base model component emulators. The grey dotted line shows the 1\% level.}
    \label{fig:MAPE}
\end{figure}

When training a NN we would like to optimise the weights $\vec{W}$ in equation \ref{eq:nonlinear_NN} such that we have sufficient prediction accuracy for all possible $\vec{\theta}$ sampled from the region of our parameter space covered by the training data. However NNs can suffer from \textit{overfitting} when trained on finite data sets (this is particularly a problem for small data sets). Overfitting occurs when the NN is learning details of the training set that are not general, resulting in high prediction accuracy when making predictions on the training set but poor accuracy when making predictions on unseen data. \textit{Ensembling} is a widely used technique in machine learning to mitigate against the problem of overfitting and improve prediction accuracy on unseen data. To train a basic ensemble of NNs, each NN is trained on the same training data but with randomly initialised weights. When predictions are made using this basic ensemble the predictions from each ensemble member are averaged. An ensemble was trained in this way in \citet{agarwal_pkann_2014} when developing an emulator for the dark matter power spectrum. We employ the same ensembling method for this work. For the ensembles that form the emulated base model components we train ensemble members until the addition of new members results is <1\% change in the prediction accuracy when making predictions on the validation set. Figure \ref{fig:MAPE} shows how the percentage change in the mean absolute percentage error (MAPE) with increasing numbers of ensemble members for the base model component emulators. The grey dotted line shows the 1\% level and the coloured dashed lines show the change in MAPE. We can see that there is some noise present in these coloured lines. This is due to the random initialisation of the NN weights, i.e., some NNs will improve the prediction accuracy more than others. To account for this we generate each ensemble multiple times and use the one that performs best on the validation set. When training the ensemble for the nonlinear boost component emulator we do not have a validation set so we can not employ the same procedure as it is important that the test set is not used at any stage of training or ensemble selection to allow us to determine the generalisation error of the emulator. From figure \ref{fig:MAPE} we can see that a majority of the improvement comes from the first $\sim 10$ ensemble members. With this in mind we train 10 ensemble members for the nonlinear boost component emulator.

All the NNs are constructed with \texttt{TensorFlow} \citep{abadi_tensorflow_2016,tensorflow_developers_tensorflow_2021}. They are all optimised with a mean squared error loss function and \texttt{Adam} optimiser \citep{kingma_adam:_2017} with $\beta_1=0.9$, $\beta_2=0.999$, $\epsilon=10^{-7}$, and a learning rate of 0.013 (the values for $\beta_1$, $\beta_2$, and $\epsilon$ are the \texttt{TensorFlow} defaults while the value for the learning rate was hand-tuned to achieve sub-percent accuracy as with the NN architectures). We train each NN for 1000 epochs or until there is a persistent (more than 20 epochs) plateau in the loss function.

%% file: sections/accuracy.tex
\subsection{Base model}
\begin{figure}
	\includegraphics[height=21cm,keepaspectratio]{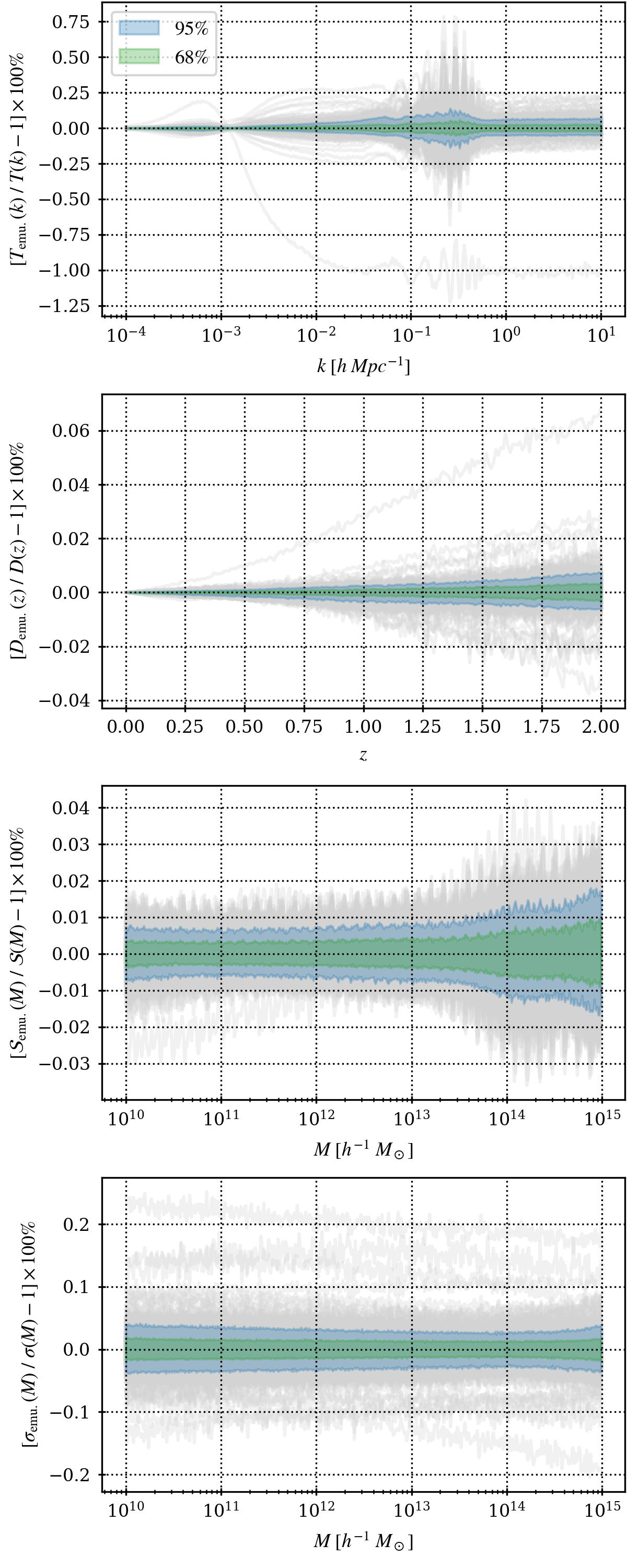}
    \caption{Percentage errors on the test set for each of the components of the base model. The panels from top to bottom correspond to the transfer function, growth factor, mass variance, and logarithmic derivative of the mass variance. The grey lines show the percentage error for each sample in the test set, the shaded regions show the 95\% and 68\% confidence intervals.}
    \label{fig:linear_accu_comp}
\end{figure}

To test the accuracy of the emulated components of the base model we use the test subsample, make predictions for each sample in the test set and compare the predictions to the results of calculating the components analytically. These comparisons are shown in figure \ref{fig:linear_accu_comp}, where we show the percentage errors for each of the components of the base model. The green and blue regions show the 68\% and 95\% confidence intervals (CIs) respectively. We can see that these shaded regions are all centred on 0\%, indicating that all component emulators produce unbiased predictions on average. We can also see that all component emulators are producing predictions with sub-percent accuracy for almost all 2000 samples in the test set (the prediction accuracy is better than 0.1\% for the 68\% CI). In the top two panels of figure \ref{fig:linear_accu_comp} there is a sample that performs considerably worse than the others. This sample represents a very extreme cosmology (with $w_0\approx-0.40$) we can see that even in this extreme case the highest level of prediction error from the transfer function component emulator is $\approx 1.25\%$. 

The top panel of figure \ref{fig:linear_accu_comp} clearly shows the highest levels of error in predictions around the BAO scale ($k \sim 0.2 \ h \ \textrm{Mpc}^{-1}$), where the BAO wiggles make these scales of $T(k)$ particularly difficult to predict. A higher level of accuracy could be achieved on these scales by increasing the number of $k$-modes around these scales, however this would mean that these scales have a greater contribution to the loss function and as such the weights of the NN would be optimised to preferentially recover these scales. For this work we aim to obtain a prediction accuracy of $<1\%$ at 68\% CI (see appendix \ref{sec:accu_reqs} on accuracy requirements) on the nonlinear galaxy power spectrum from \texttt{matryoshka}. As is shown in section \ref{sec:accuracy} this is achievable without exploring this solution and therefore we leave it to future works.

\begin{figure*}
	\includegraphics[width=\linewidth]{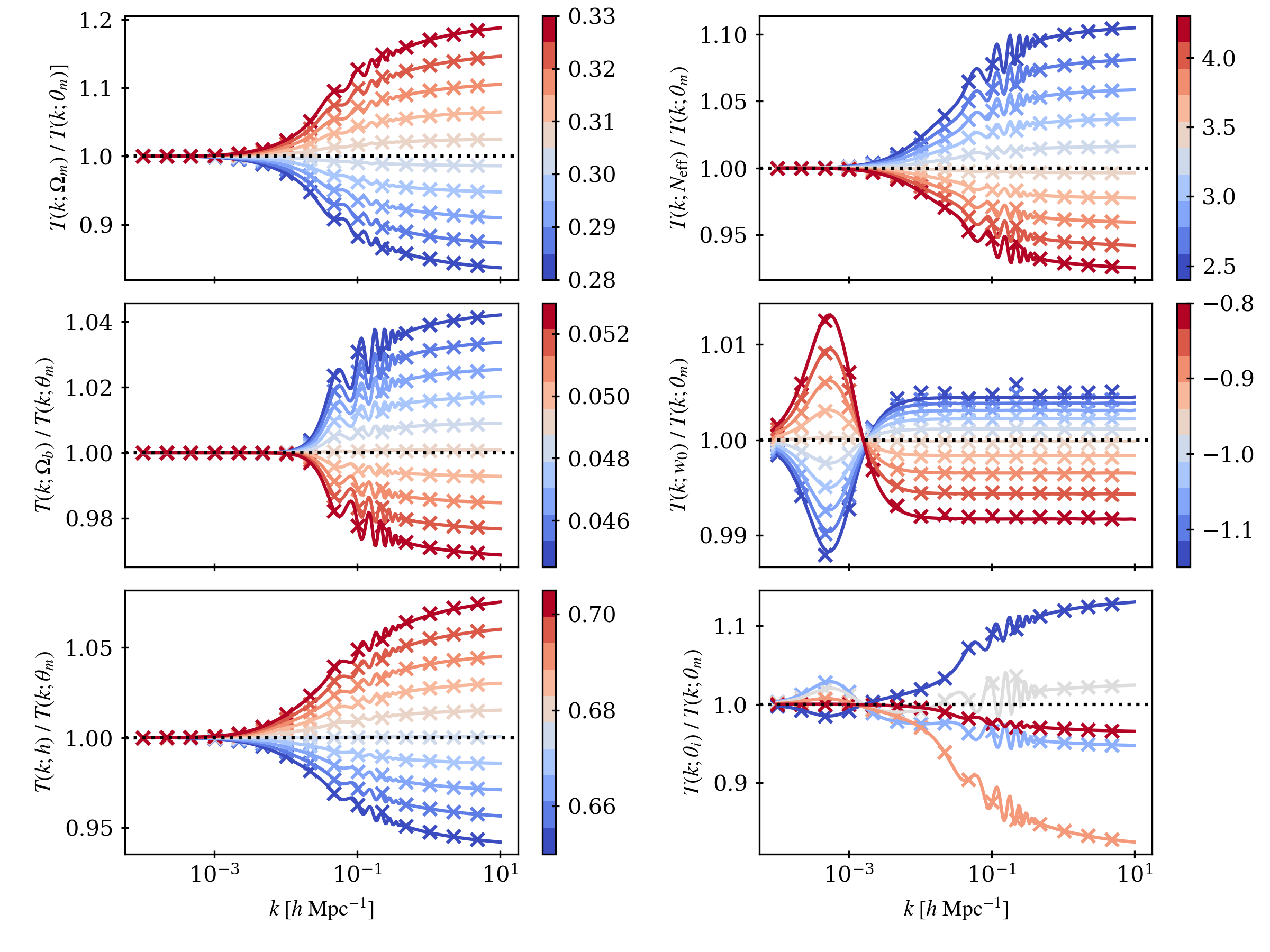}
    \caption{Plots showing the response of the transfer function to relevant parameters of our cosmological model. All panels except the bottom right show the effect of varying a single parameter of the model, the bottom right panel shows five random parameter sets. In all panels the ratio is with respect to the transfer function calculated for parameters that correspond to the mean of the training space shown in figure \ref{fig:linear_emu_space}. The solid lines show responses of transfer functions calculated with \texttt{CLASS}, the crosses show the emulator predictions for the same parameters. In all except the lower right panel the lines and crosses are coloured by the value of the parameter that is being varied, in the lower right panel the colour corresponds to the index of the test set from which the parameter sets were randomly selected.}
    \label{fig:transfer_grad}
\end{figure*}

Figure \ref{fig:transfer_grad} shows the response of the transfer function to various parameters of our cosmological model, when calculating the transfer function with \texttt{CLASS} and making a prediction for the transfer function with the emulator. Figure \ref{fig:transfer_grad} shows that the emulator well recovers the response of the transfer function for all the relevant parameters of our model, however there is some indication of a slight bias in the prediction of the transfer function on small scales for increasingly negative values of $w_0$. This bias is $\ll 1\%$ and only occurs for extreme values of $w_0$, so we don't expect this small bias to have a significant impact when using the emulator to make predictions for less extreme cosmologies.

\subsection{Nonlinear Boost}

\begin{figure}
	\includegraphics[width=\columnwidth]{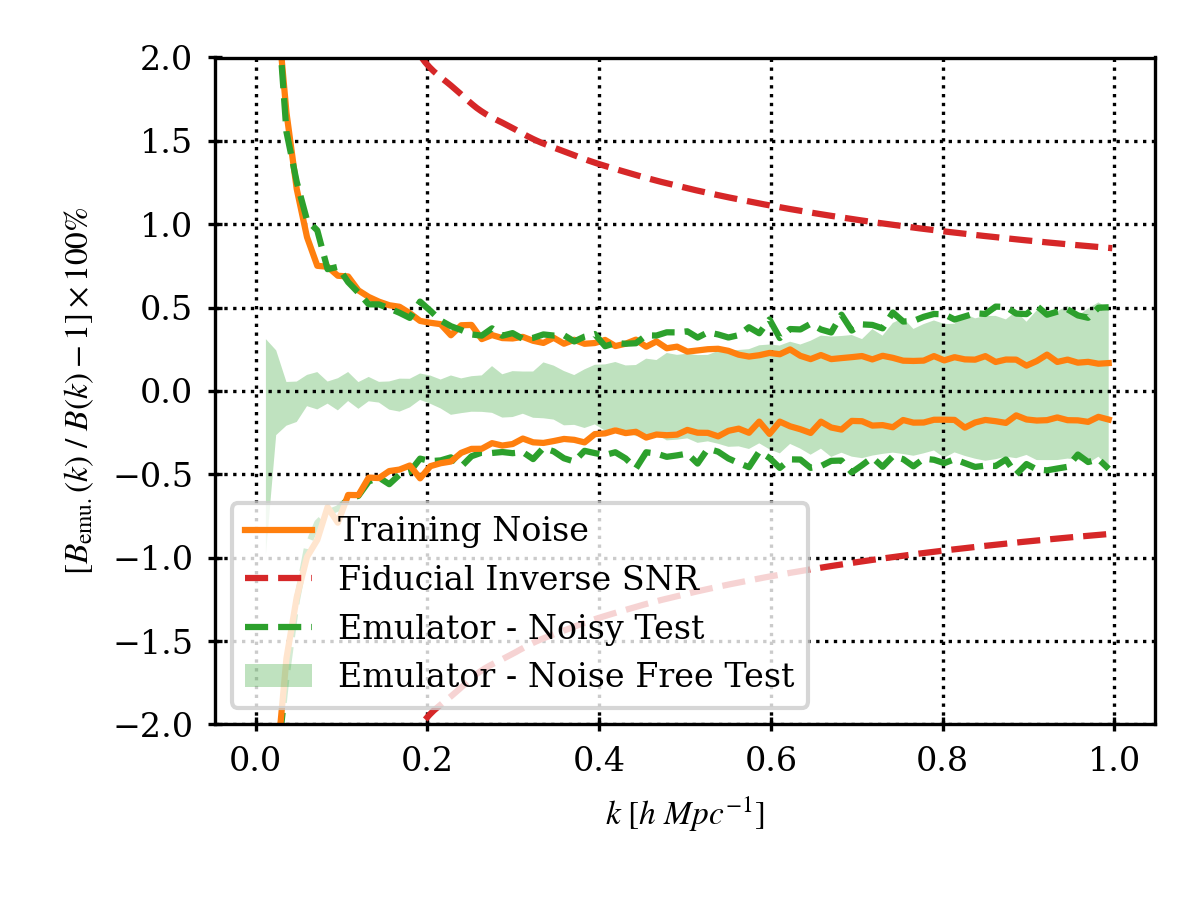}
    \caption{Prediction error from the nonlinear boost component emulator compared to the statistical error of our fiducial mock and the noise level of the training data. The green shaded region represents the true $1\sigma$ prediction error of the emulator, the green dashed line represents the $1\sigma$ prediction error measured when evaluating the emulator predictions with a noisy test data. The orange solid line shows the $1\sigma$ noise level in the nonlinear boost training data, and the red dashed line shows the inverse signal to noise ratio (SNR) of our fiducial mock observation in percent.}
    \label{fig:boost_accu}
\end{figure}

To test the performance of the nonlinear boost emulator we calculate nonlinear boosts for the seven Aemulus test cosmologies. As with the training data we generate power spectra for 50 different HOD parameter sets for each cosmology, which results in 350 test samples. We generate two versions of this test set: the first contains the same level of noise as the training data and the second contains no noise. The reasons for generating these two versions of the same test set is clear when looking at figure \ref{fig:boost_accu}, which shows percentage error on the predictions from the nonlinear boost component emulator. The green shaded region shows the prediction error (68\% CI) on the noise free test set and the green dashed line shows the prediction error on the noisy test set. We can see that for $k \gtrsim 0.8 \ h \ \mathrm{Mpc}^{-1}$ the green dashed line and shaded region agree while on larger scales we can see that the prediction error calculated using the noisy test set very closely follows the noise level of the noisy test set. This indicates that the calculated prediction error is dominated by the noise on these scales, which is confirmed by looking at the prediction error on the noise free test set on the same scales. The NNs are able to predict the nonlinear boost at higher accuracy than the noise level of the training data because the noise is random and therefore the noise averages across the training set. When using a simulation suite it is not possible to generate a noise free test set. The test simulations in the Aemulus suite do however have multiple realisations, such that the noise level in the test set generated from these simulations is lower than that of the training set allowing for an accurate assessment of the prediction accuracy.

From figure \ref{fig:boost_accu} we can see that the nonlinear boost component emulator has $\lesssim 0.5\%$ prediction error from $0.01 \ h \ \mathrm{Mpc}^{-1} \lesssim k \lesssim 1 \ h \ \mathrm{Mpc}^{-1}$. The level of prediction error is largest on the smallest scales. This is where the dynamic range of the nonlinear boost is largest. This prediction error gets smaller going to larger scales, however there is a spike in prediction error at $k \sim 0.1 \ h \ \mathrm{Mpc}^{-1}$. This spike in prediction error is a result of the noise level of the training set having an impact on the NNs ability to learn the nonlinear boost on these scales. To reduce the impact of this when producing predictions for the nonlinear galaxy power spectrum (by combing the predictions of the boost with the base model prediction), the predictions of the nonlinear boost can be smoothly "stitched" with those that we expect from linear theory (that being $B(k)\approx1$ on large scales) as was done in \citet{kobayashi_accurate_2020}. We are able to produce predictions for the nonlinear galaxy power spectrum with \texttt{matryoshka} with <1\% error (at 68\% CI) without exploring this solution, as such we leave this for future works.

\begin{figure}
	\includegraphics[width=\columnwidth]{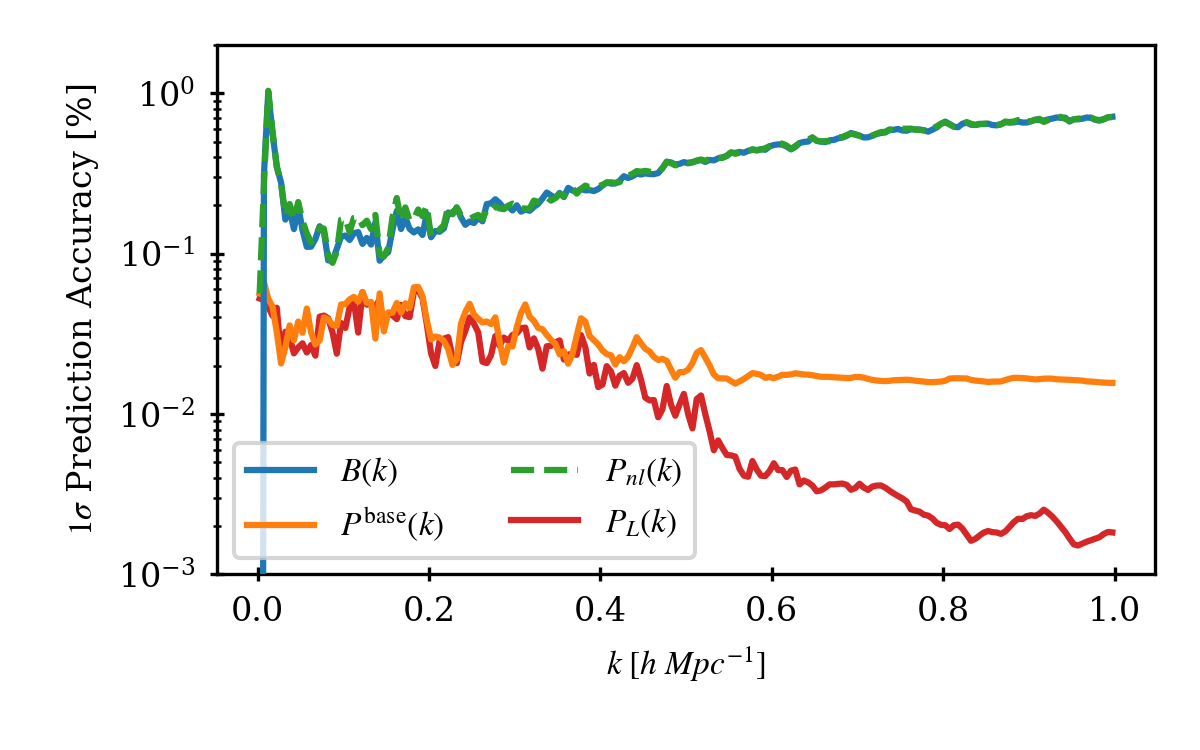}
    \caption{Plot showing the relative contributions to the prediction error on the nonlinear galaxy power spectrum $P_{nl}(k)$. The prediction error from each of the component emulators contributes to the overall prediction error on $P_{nl}(k)$. For simplicity we have only shown the error coming from the nonlinear boost $B(k)$ component emulator, the error in the base model $P^\mathrm{base}(k)$ (which is calculated with predictions from all the base model component emulators), and the error in the prediction of the linear matter power spectrum $P_L(k)$ (which is calculated with predictions from the transfer function component emulator).}
    \label{fig:relative_err}
\end{figure}

We can examine the relative contributions from the emulated base model and the nonlinear boost component emulator to the prediction error on the nonlinear galaxy power spectrum. Figure \ref{fig:relative_err} compares the $1\sigma$ prediction errors on the nonlinear galaxy power spectrum to those on: the nonlinear boost, the emulated base model (which is calculated with predictions from all the base model component emulators), and the linear matter power spectrum (which is calculated with predictions from the transfer function component emulator). We can immediately see that the prediction error from the nonlinear boost is dominating on all scales. We can also see that on small scales the contribution to the error from the linear power spectrum (and thus the transfer function) is lower than the emulated base model. This is to be expected as on these small scales the 1-halo term (equation \ref{eq:p_1h}) is dominating, and the prediction errors from the other base model component errors are more significant.

The prediction accuracy of the nonlinear boost component emulator (the dominating component of the \texttt{matryoshka} prediction error) is compared to the inverse signal to noise ration of our fiducial mock in figure \ref{fig:boost_accu} (with a volume of $(1 \ \mathrm{Gpc} \ h^{-1})^3$ and number density of $\sim 6\times 10^{-4} \ (\mathrm{Mpc}^{-1} \ h)^3$). We can see that the prediction error from the nonlinear boost component emulator is significantly lower than the statistical error of our fiducial mock on all scales considered. This implies that the achieved level of prediction accuracy is high enough to produce predictions for the power spectrum that are consistent with the truth within the statistical errors of our mock analyses setup (see appendix \ref{sec:accu_reqs} for a discussion on how the required prediction accuracy depends on sample considered). 

\begin{figure*}
	\includegraphics[width=\linewidth]{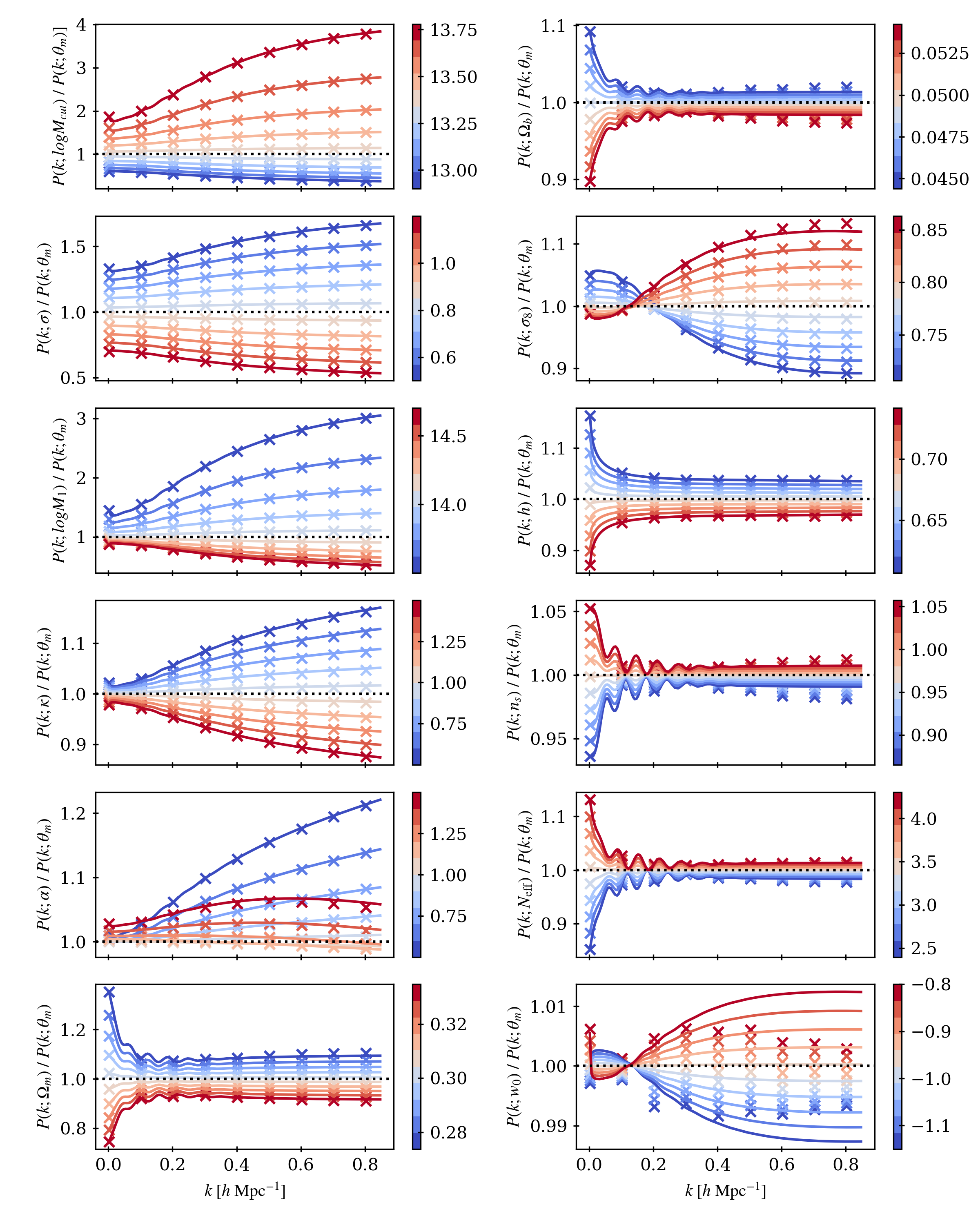}
    \caption{Similar to figure \ref{fig:transfer_grad}, for the response of the nonlinear galaxy power spectrum to the cosmological and HOD parameters of our model.}
    \label{fig:nonlinear_grad}
\end{figure*}

Figure \ref{fig:nonlinear_grad} shows the response of the nonlinear galaxy power spectrum to the cosmological and HOD parameters of our model. We can see that the response is generally well recovered by \texttt{matryoshka}, particularly on large scales, however the response is not well recovered on small scales for all parameters. This is most apparent in the response to $w_0$, where we can see that the response is under predicted on small scales. This effect is greater the further the value of $w_0$ deviates from -1. It should be noted that the response of the nonlinear galaxy power spectrum to $w_0$ is very small. The difference between the power spectra for $w_0=-1.15$ and $w_0=-0.8$ is $\sim 2\%$. This response is considerably smaller than any of the other parameters. The result of this is that the response to $w_0$ has the least impact on the loss function. This effect is exaggerated by the noise in the training data for the nonlinear boost component emulator. The $1\sigma$ noise level at $k=0.8 \ h \ \mathrm{Mpc}^{-1}$ is $\sim 0.2\%$, which is similar to the the difference between power spectra calculated with $w_0=-1.15$ and $w_0=-1.11$.

%% file: sections/fullshape.tex
In this section we test \texttt{matryoshka} by doing several FS analyses of a mock observed power spectrum. These analyses are summarised in table \ref{tab:setups}. These mock analyses aim to verify that we can recover unbiased constraints on the input cosmology when conducting a FS analysis with \texttt{matryoshka}. They also allow us to investigate the potential gain in constraining power by including smaller scales when conducting a FS analysis of the power spectrum.

\begin{table*}
 \centering
 \begin{tabular}{c c p{0.6\linewidth}} 
 \hline
 Setup & $k_\mathrm{max} \ [h \ \mathrm{Mpc}^{-1}]$ & Notes \\ 
 \hline
 Fiducial & 0.25 & Our fiducial full shape analysis. Designed to mimic the scales used in an EFTofLSS style analysis.\\
 \hline
 Including Number Density & 0.25 & Repeats of our fiducial analysis with the inclusion of the number density in the likelihood as in equation \ref{eq:likelihood_nd}.\\
 \hline
 Small Scales & [0.45-0.85] & Repeats of our fiducial analysis with increased values of $k_\mathrm{max}$.\\
 \hline
 Fixed HOD & [0.25-0.85] & Repeats of our fiducial analysis with all HOD parameters fixed to the truth, and with the increased values of $k_\mathrm{max}$ of the small scales analyses.\\
 \hline
 \end{tabular}
\caption{Table summarising the different full shape analyses described in section \ref{sec:fullshape}. Unless otherwise stated all analysis setups have $k_\mathrm{max}=0.025 \ [h \ \mathrm{Mpc}^{-1}]$.}
\label{tab:setups}
\end{table*}

\subsection{Mock Observation}
\label{subsec:mock_cmass}

We produce a mock observed power spectrum designed to approximate the power spectrum of BOSS CMASS galaxies, with nonlinearities coming from HALOFIT as with the nonlinear training data (see ection \ref{subsubsec:fake_sim}). The cosmological and HOD parameters corresponding to this mock observation are shown in tables \ref{tab:linear_param} and \ref{tab:HOD_param} respectively. The cosmological parameters come from the most recent Planck $\Lambda$CDM TT, TE, EE + lowE + lensing + BAO analysis \citep[table 2 in][henceforth Planck 2018]{planck_collaboration_planck_2020}. The HOD parameters are the best fit parameters that result from the small scale clustering analysis of BOSS CMASS galaxies conducted by \citet{white_clustering_2011}. The number density associated to this mock observation is $\sim 6\times 10^{-4} \ (\mathrm{Mpc}^{-1} \ h)^3$. It should be noted that this number density is slightly greater than the observed CMASS number density. This value corresponds to the number density calculated using the Planck 2018 cosmological parameters and \citet{white_clustering_2011} best fit HOD parameters with the equation 
\begin{equation}
    n_g = \int n(M)\langle N | M \rangle dM\ .
    \label{eq:number_dens}
\end{equation}
We include linearly spaced $k$-bins covering the range $0.0025 \ h \ \textrm{Mpc}^{-1} < k < 0.85 \ h \ \textrm{Mpc}^{-1}$ with $\Delta k = 0.005 \ h \ \textrm{Mpc}^{-1}$. These scales are selected such that the $k$-bins included in our fiducial analysis (with $k_\textrm{max} = 0.25 \ h \ \textrm{Mpc}^{-1}$) match those from \citet{ivanov_cosmological_2020}, which used the perturbation theory based EFTofLSS approach to analyse multipoles of the power spectra of BOSS galaxies.

\begin{figure}
	\includegraphics[width=0.95\columnwidth]{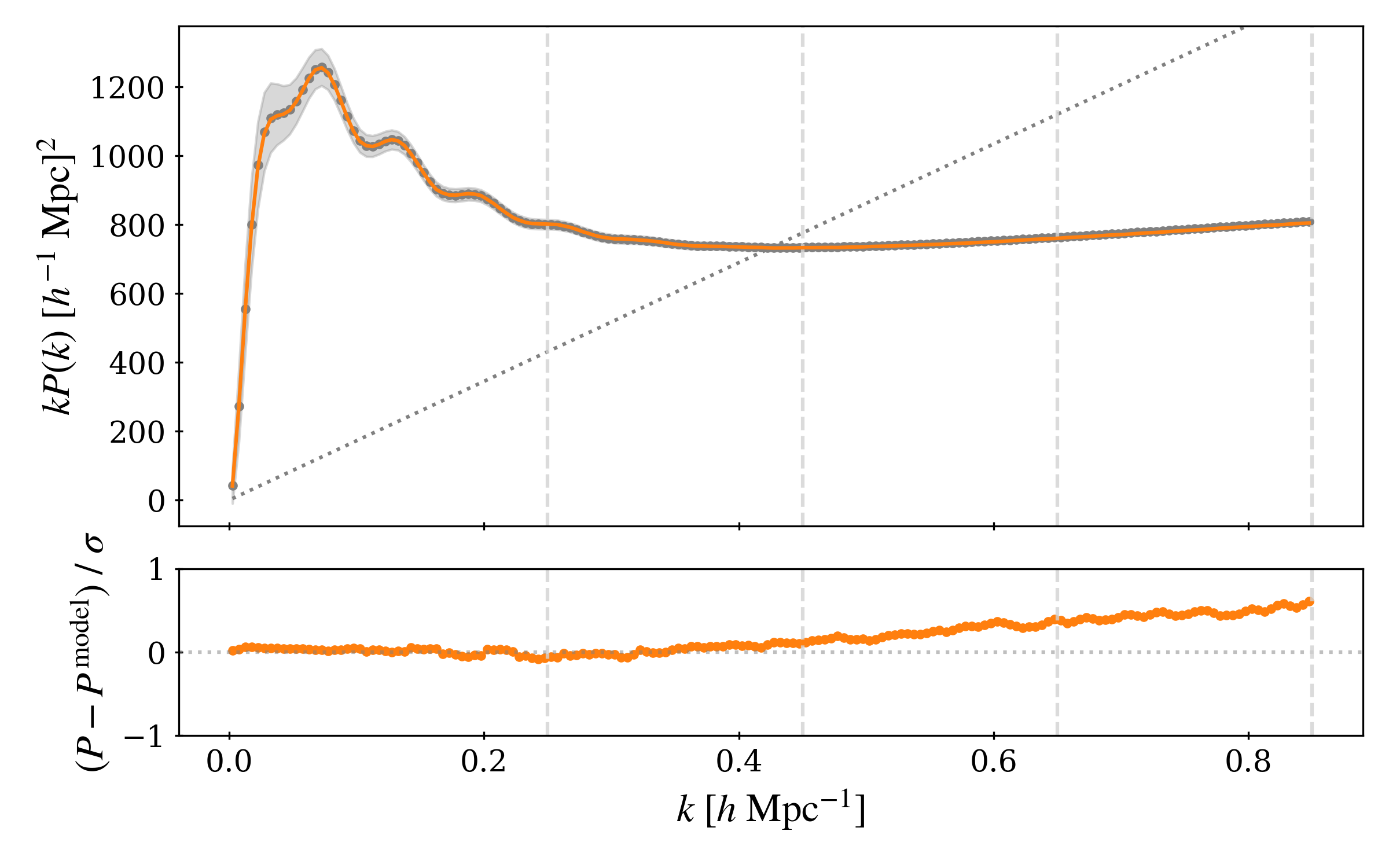}
    \caption{The top panel shows the mock CMASS power spectrum described in section \ref{subsec:mock_cmass} (grey points and shaded region), as well as the \texttt{matryoshka} prediction for the fiducial parameters (solid orange line). The vertical dashed lines show the $k_\mathrm{max}$ values of the full shape analyses. The dotted line shows the shot noise level of our mock observation. The bottom panel shows the normalised residuals comparing our mock observation to the \texttt{matryoshka} prediction.}
    \label{fig:mock_observation}
\end{figure}

Figure \ref{fig:mock_observation} shows our mock observation with grey points. We calculate an uncertainty for this mock observation using equation \ref{eq:gauss_cov} with a volume of $(1 \ \mathrm{Gpc} \ h^{-1})^3$. This uncertainty is shown with the grey shaded region in figure \ref{fig:mock_observation}. The \texttt{matryoshka} prediction for the fiducial parameters is shown with the orange solid line. We can see that the \texttt{matryoshka} prediction only becomes distinguishable from the truth at very small scales ($k\gtrsim 0.5 \ h \mathrm{Mpc}^{-1}$), but is still consistent with the truth at the $1\sigma$ level. The FS analyses that follow will determine to what extent this small error in the prediction of the power spectrum on small scales impacts the constrained cosmology.\\

\subsection{Fiducial Analysis}
\label{subsec:fiducial}

For our fiducial analysis we fit four out of the five HOD parameters of our model ($\log{M_\mathrm{cut}}, \sigma, \log{M_1}, \alpha$) and five out of the seven cosmological parameters ($\Omega_m, \sigma_8, h, n_s, w_0$). We fix $\kappa$ to it's true value as it is not well constrained by the real space power spectrum for the scales considered in our fiducial analysis (or any of the FS analyses that follow). The purpose of the analyses of this section is to verify that unbiased cosmology can be recovered, as such fixing $\kappa$ to the truth in this way doesn't influence any conclusions we draw. We also fix $\Omega_b$ and $N_\mathrm{eff}$ to their true values. We do not expect to get competitive constraints on these parameters from the real space power spectrum. It is common practice to use a very tight prior on $\Omega_b$ informed by big bang nucleosynthesis, and to fix $N_\mathrm{eff}=3.046$ to align with standard model predictions.

We use Markov Chain Monte Carlo (MCMC) sampling to calculate the posterior distributions of HOD and cosmological parameters. We define a Gaussian likelihood with the form
\begin{equation}
    \ln{\mathcal{L}(d|\theta,\phi)} = -\frac{1}{2}(P-\tilde{P})^T\mathbfit{C}^{-1}(P-\tilde{P})\, ,
    \label{eq:likelihood}
\end{equation}
where $P$ is the mock observed galaxy power spectrum, $\mathbfit{C}$ is the Gaussian covaraince matrix calculated using equation \ref{eq:gauss_cov} shown by the grey shaded region in figure \ref{fig:mock_observation}, and $\tilde{P}$ is the emulated galaxy power spectrum. We do not include any information about the galaxy number density in the likelihood for our fiducial analysis, however the number density is very sensitive to the HOD parameters. We explore the impact of including number density information in the likelihood in section \ref{subsec:number_dens}.

We use flat priors on all of the free HOD parameters with ranges equivalent to the extent of the HOD training space (see table \ref{tab:HOD_param}). For the cosmological parameters we use a multivariate Gaussian prior with mean and covariance defined by the training samples for the emulated base model shown in figure \ref{fig:linear_emu_space}. This is a very wide prior, as mentioned in section \ref{subsubsec:analytic_data}, which covers a $4\sigma$ region coming from previous CMB, BAO, and supernovae analyses. The use of this multivariate Gaussian prior on the cosmological parameters is necessary to assign low probability to areas of the parameter space that have not been sampled with training data, as the predictions from the emulators will not be accurate in these regions of the parameter space. MCMC sampling is done using \texttt{zeus} \citep*{karamanis_zeus_2021}, \texttt{zeus} uses ensemble slice sampling, a method that is robust when sampling from challenging distributions which is often the case for HOD parameters. Convergence of MCMC chains is discussed in appendix \ref{sec:convergence}.

\begin{figure*}
	\includegraphics[width=\linewidth]{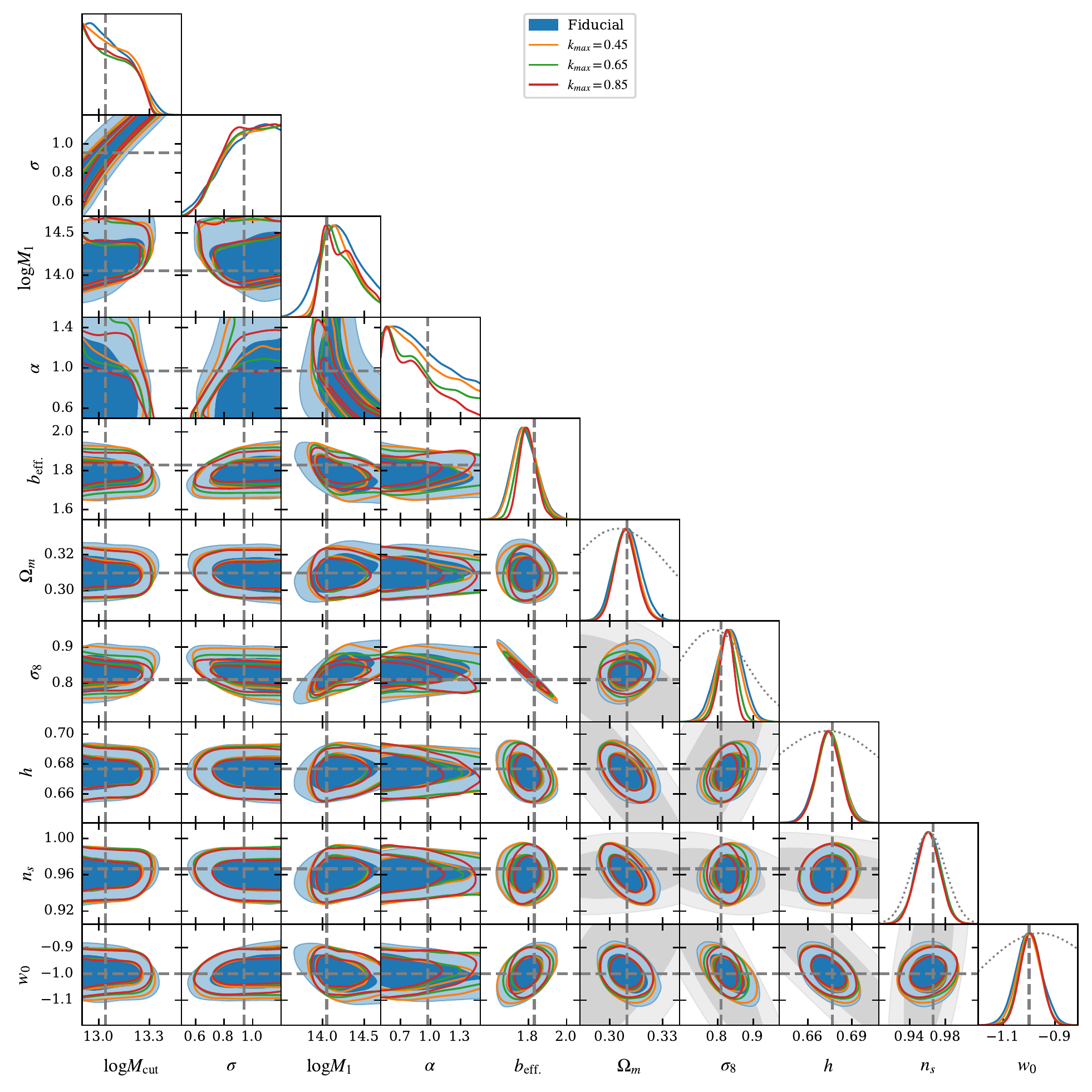}
    \caption{Marginalised 1D and 2D posterior distributions resulting from the full shape analyses described in sections \ref{subsec:fiducial} and \ref{subsec:increase_kmax}. The two contour levels in the off-diagonal panels represent the $1\sigma$ and $2\sigma$ regions. The grey dashed horizontal and vertical lines show the true cosmological and HOD parameters. The grey contours in the off-diagonal panels and grey dotted lines in the diagonal panels show the multivariate prior on the cosmological parameters. All model parameters not shown are fixed to the truth.}
    \label{fig:kmax_corner}
\end{figure*}

The posterior distributions calculated from our fiducial analysis are shown in figure \ref{fig:kmax_corner} with blue filled contours. We can see that the true cosmological parameters are recovered within $1\sigma$, verifying that the obtained level of prediction accuracy from \texttt{matryoshka} is sufficient to return unbiased cosmology for our mock. We also show the marginalised posterior distributions for the effective galaxy bias $b_\mathrm{eff.}$ in figure \ref{fig:kmax_corner}. $b_\mathrm{eff.}$ is not a free parameter but can be calculated from the cosmological and HOD parameters with the equation
\begin{equation}
    b_\mathrm{eff.} = \frac{1}{n_g}\int n(M)b_h(M)\langle N|M \rangle  \ .
\end{equation}
We can see there is a strong, and expected, degeneracy between $b_\mathrm{eff.}$ and cosmological parameters that primarily impact the amplitude of the power spectrum such as $\sigma_8$. The effective bias is sensitive to the HOD parameters, which are more tightly constrained by small scales. Therefore we expect to see an improved constraint on the cosmological parameters when including smaller scales. This improvement is coming from the increase in statistical power, along with the improved constraint on the HOD parameters, and thus the effective bias.

\subsection{Impact of Number Density}
\label{subsec:number_dens}

In our fiducial analysis we do not include any information about the number density in our likelihood. The number density is sensitive to the HOD parameters. Works such has \citet{zhou_clustering_2020} and \citet{lange_five-percent_2021} have included information about the number density via an extra term in the likelihood, such that
\begin{equation}
    \ln{\mathcal{L}(d|\theta,\phi)} = -\frac{1}{2} \left[ (P-\tilde{P})^T\mathbfit{C}^{-1}(P-\tilde{P})+\frac{(n_g - \tilde{n_g})^2}{\sigma_{n_g}^2}\right]\ ,
    \label{eq:likelihood_nd}
\end{equation}
where $P$, $\tilde{P}$, and $\mathbfit{C}$ are the same as in equation \ref{eq:likelihood}, $n_g$ is the observed number density, $\tilde{n}_g$ is the number density predicted by the model which can be calculated via equation \ref{eq:number_dens}, and $\sigma_{n_g}$ is the uncertainty on the observed number density. \citet{miyatake_cosmological_2020} noted very little change to inferred cosmological parameters by including information about the number density when analysing projected clustering and weak lensing of a CMASS like sample. To investigate the impact of the number density when doing a FS analysis of the power spectrum, we re-run our fiducial analysis with $\sigma_{n_g}=[0.1,0.05,0.01]n_g$. The decreasing values of $\sigma_{n_g}$ have a similar impact to placing tighter priors on the HOD parameters.

\begin{figure}
	\includegraphics[width=\linewidth]{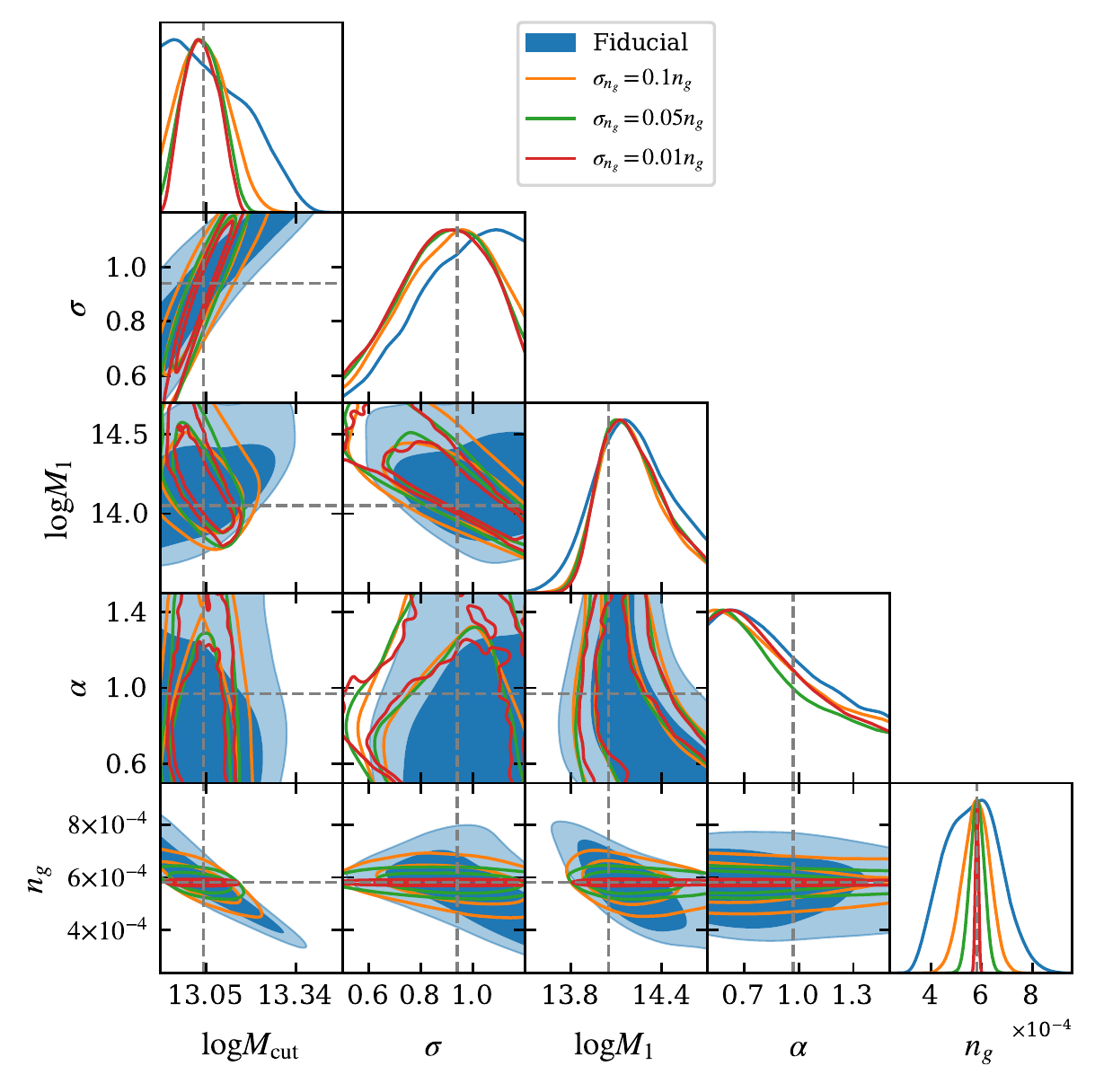}
    \caption{Similar to figure \ref{fig:kmax_corner}. The blue filled contours are the same as in figure \ref{fig:kmax_corner}. The empty contours show the results of the number density analyses described in section \ref{subsec:number_dens}. Each of the empty contours shows the results of a full shape analysis over the same scales of the fiducial analysis with different levels of assumed uncertainty on the number density from the mock. Only the HOD parameter posteriors are shown here as the impact on the cosmological parameters is negligible, see figure \ref{fig:setup_comp_1d} for the 1D cosmological posteriors.}
    \label{fig:corner_number_dens}
\end{figure}

\begin{figure*}
	\includegraphics[width=\linewidth]{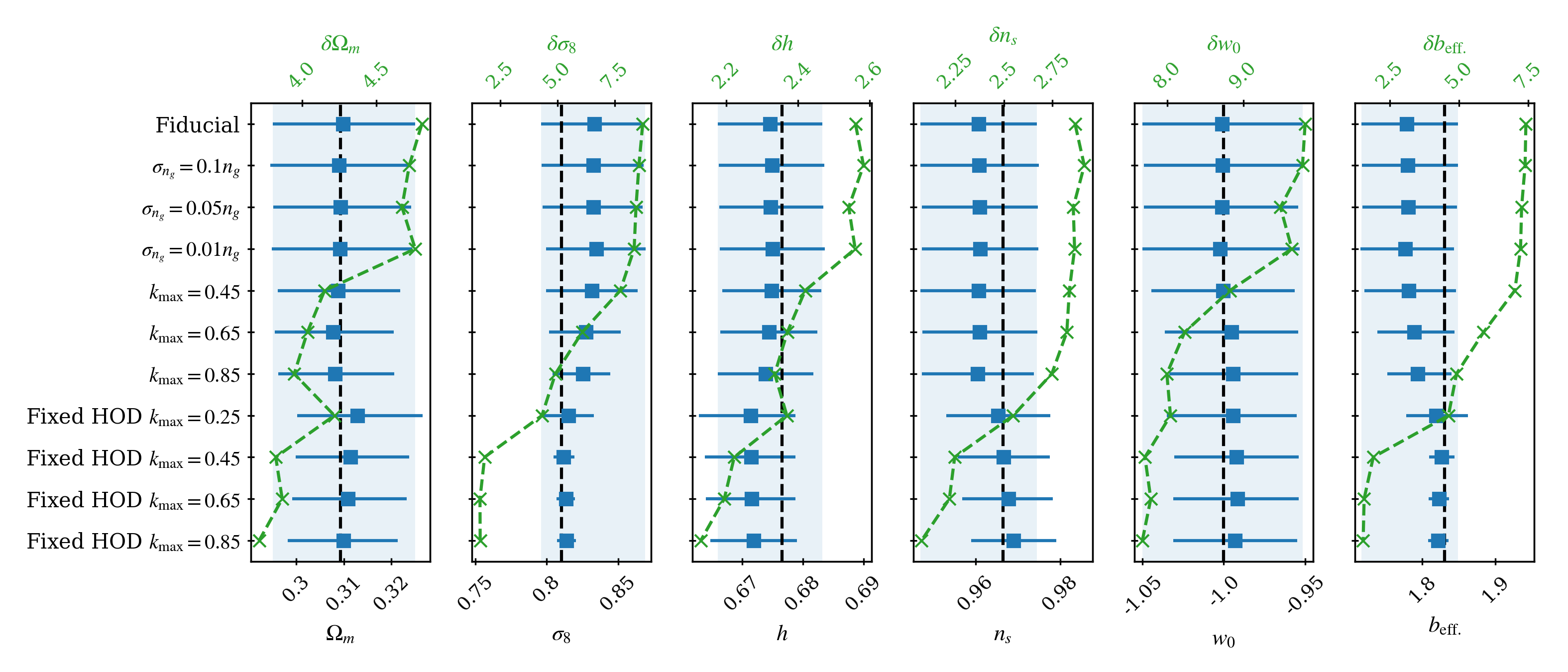}
    \caption{Percent constraint and marginalised 1D posteriors for each of the cosmological parameters considered in the full shape analyses of section \ref{sec:fullshape}. The green crosses and green dashed line shows the percent constrain and corresponds to the top x-axis. The blue points and error bars show the median and $1\sigma$ region of the marginalised posteriors and correspond to the bottom x-axis. The blue shaded region shows the width of our fiducial analysis and the black vertical dashed line shows the location of the truth.}
    \label{fig:setup_comp_1d}
\end{figure*}

The posterior distributions of the HOD parameters resulting from these analyses are shown in figure \ref{fig:corner_number_dens}, alongside the results from the fiducial analysis for comparison. The cosmological parameters are not shown as the difference between the inferred cosmology from these analyses is minimal (the marginalised 1D posteriors for the cosmological parameters are shown in figure \ref{fig:setup_comp_1d} for reference), however we do show the marginalised posteriors for the number density predicted by our model. We can see that even for our fiducial analysis the true value of $n_g$ is well recovered. We can also see that even a relatively high value of $\sigma_{n_g}$ significantly improves the constraint on some of the HOD parameters, particularly $\log{M_\mathrm{cut}}$. This is to be expected as $\log{M_\mathrm{cut}}$ directly controls the number of central galaxies (which make up the majority of a CMASS-like sample), and thus the number density.

\subsection{Increasing $k_\mathrm{max}$}
\label{subsec:increase_kmax}

To investigate the impact of the minimum scale included when conducting a FS analysis of the power spectrum on the constraint on the cosmological and HOD parameters, we re-run our fiducial analysis pushing the value of $k_\mathrm{max}$ to $[0.45, 0.65, 0.85] \ h \ \mathrm{Mpc}^{-1}$. The 1D and 2D marginalised posteriors resulting from these analyses are shown in figures \ref{fig:kmax_corner} and \ref{fig:setup_comp_1d}. As expected we see significant improvement in the constraint on all cosmological parameters by including smaller scales. This is particularly true for $\Omega_m$ and $\sigma_8$. Our fiducial analysis results in a $\sim 8.7\%$ ($\sim 4.8\%$) constraint on $\sigma_8$ ($\Omega_m$). Pushing the minimum scale to $k_\mathrm{max}=0.85 \ h \ \mathrm{Mpc}^{-1}$ results in a $\sim 4.9\%$ ($\sim 3.9\%$) constraint on $\sigma_8$ ($\Omega_m$), which represents a $\sim 1.8 \times$ ($\sim 1.2 \times$) improvement. This improvement is coming from the higher statistical power of smaller scales in addition to the improved constraint on the HOD parameters that arises from the increasing magnitude of the response of the power spectrum to the HOD parameters shown in figure \ref{fig:nonlinear_grad}.

We can see from figures \ref{fig:kmax_corner} and \ref{fig:setup_comp_1d} that there is a small ($< 1\sigma$) bias in the median values (blue squares in figure \ref{fig:setup_comp_1d}) of the marginalised posteriors from these analyses. The origin of this bias is likely the small error in the emulator predictions of the power spectrum. This effect is exaggerated by degeneracies between model parameters. For example the median value for $\sigma_8$ is over-predicted whilst $b_\mathrm{eff.}$ is under-predicted. Improving the constraint on $b_\mathrm{eff.}$ by including smaller scales (or fixing the HOD as described in section \ref{subsec:fixed_hod}) reduces the observed bias in $b_\mathrm{eff.}$ and $\sigma_8$.

\subsection{Fixed HOD}
\label{subsec:fixed_hod}

\begin{figure}
	\includegraphics[width=\linewidth]{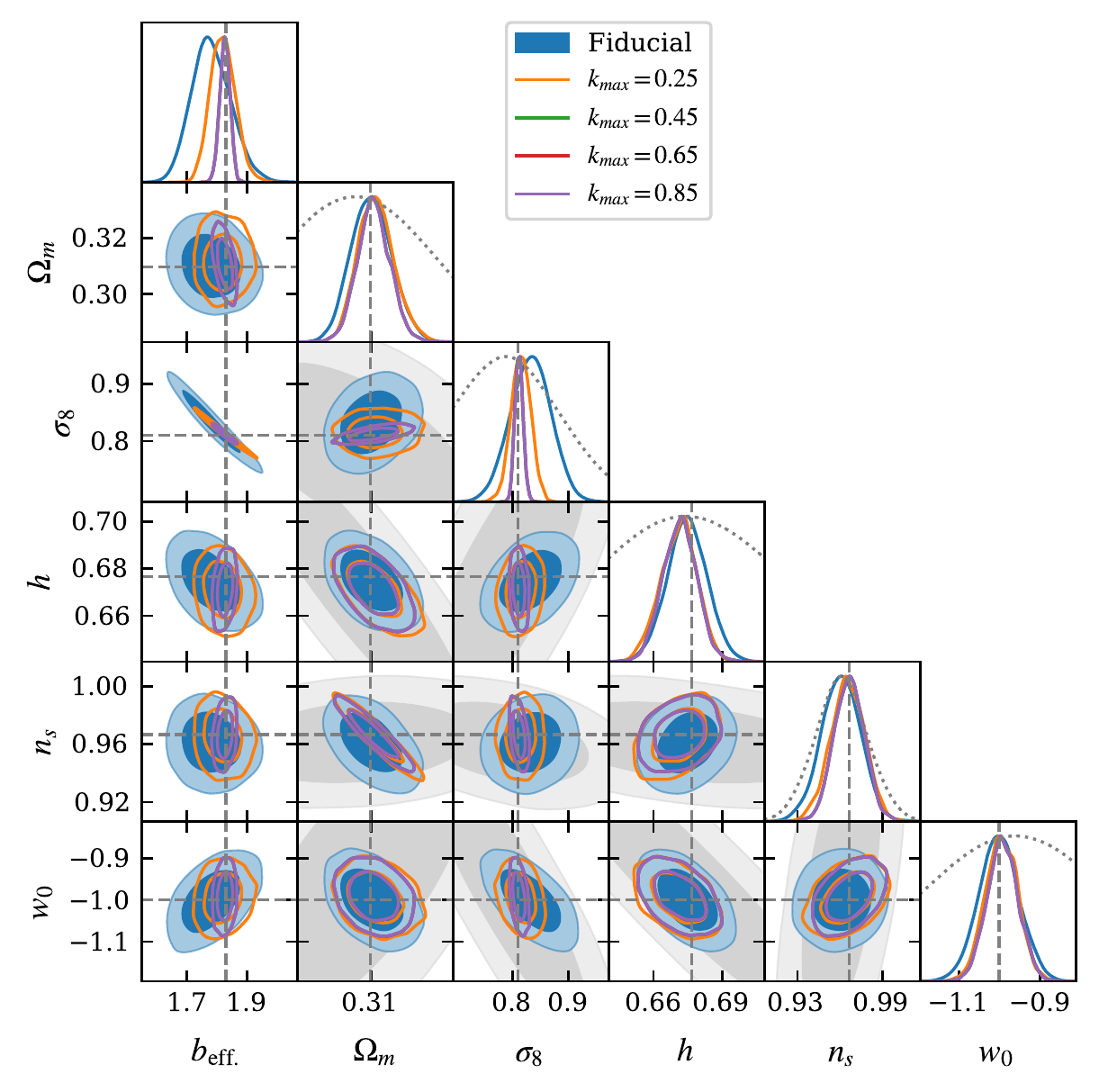}
    \caption{Similar to figure \ref{fig:kmax_corner}. The blue filled contours are the same as in figure \ref{fig:kmax_corner}. The empty contours show the results of the fixed HOD analyses described in section \ref{subsec:fixed_hod}, where only the five cosmological parameters shown are allowed to vary (all other model parameters are fixed to the truth).}
    \label{fig:fixHOD_corner}
\end{figure}

To investigate to what level the cosmological parameter constraints are degraded by fitting the HOD parameters, we re-run our fiducial analysis with all HOD parameters fixed to the truth, and for the same $k_\mathrm{max}$ values used in section \ref{subsec:increase_kmax}. The results of these fixed HOD analyses are shown in figures \ref{fig:setup_comp_1d} and \ref{fig:fixHOD_corner}. The orange contours show the results that cover the same scales as our fiducial analysis. We can see that compared to the results from the fiducial analysis (blue filled contours) there is a significant increase in the constraint on all cosmological parameters even when the same scales are considered. The improvement on the constraint on $\sigma_8$ is $\sim 2.0\times$, which is already larger than the improvement from pushing to $k_\mathrm{max}=0.85 \ h \ \mathrm{Mpc}^{-1}$ in section \ref{subsec:increase_kmax}. This result is expected and demonstrates just how much fitting the HOD parameters degrades the constraint on the cosmological parameters, and highlights that accurate prior information about the HOD coming from small scale clustering studies \citep[such as][]{zheng_galaxy_2007,white_clustering_2011,parejko_clustering_2013,beutler_6df_2013,zhai_clustering_2017} can greatly improve constraints on cosmology coming from a full shape analysis of the power spectrum in the HM framework. We can also see that increasing $k_\mathrm{max}$ compared to our fiducial analysis results in further improvement to the constraint on cosmology. Pushing the fixed HOD analysis to $k_\mathrm{max}=0.45$ results in a $\sim 4.8\times$ improvement to the constraint on $\sigma_8$, however we notice that the improvement including scales smaller than this is much less significant. The dotted line in figure \ref{fig:mock_observation} shows the shot noise of our mock observation. We can see that when $k \sim 0.4 \ h \ \mathrm{Mpc}^{-1}$ the shot noise is the same magnitude as our clustering signal. As such, increasing $k_\mathrm{max}$ to values $\lesssim 0.4 \ h \ \mathrm{Mpc}^{-1}$ will result in a higher signal to noise ratio, however increasing $k_\mathrm{max}$ to values $\lesssim 0.4 \ h \ \mathrm{Mpc}^{-1}$ will not. The reason we see continued gain when pushing $k_\mathrm{max}\gtrsim 0.4 \ h \ \mathrm{Mpc}^{-1}$ for the analyses of section \ref{subsec:increase_kmax} but not these fixed HOD analyses is due to the scale dependence of the response to the HOD parameters shown in figure \ref{fig:nonlinear_grad}. We can see that increasing $k_\mathrm{max}$ increases sensitivity to all the HOD parameters, however for the cosmological parameters there is no scale dependence beyond $k_\mathrm{max}\gtrsim 0.4 \ h \ \mathrm{Mpc}^{-1}$.

%% file: sections/disc.tex
\subsection{Inclusion of Additional Effects Through B(k)}
\label{subsec:boost_absorb}
The base model of \texttt{matryoshka} uses fitting functions for quantities such as the halo mass function and the concentration-mass relation. Furthermore, for this work we are ignoring redshift space distortions and effects such as halo exclusion. The base model of \texttt{matryoshka} has been designed to be used alongside simulations. Effects that cause $B(k)$ to differ significantly from unity on large scales will need to be included in the base model in future work. On the other hand, effects that predominately impact small scales (that can be challenging to model analytically in some cases) do not to necessarily be included as they can effectively be absorbed into the prediction of $B(k)$. For example, redshift space distortions (RSD) can be modelled on large scales with a Kaiser factor
\begin{equation}
    P_s^\mathrm{base}(k, \mu) = P(k)(1+f\mu^2)^2\ , 
\end{equation}
while the impact of RSD on small scales is more challenging to model analytically. To make sure the small scale RSD is included in the \texttt{matryoshka} predictions without modelling them analytically we can decompose $P_s^\mathrm{base}(k, \mu)$ into multipoles
\begin{equation}
    P_\ell^\mathrm{base}(k) = \frac{2\ell+1}{2}\int^1_{-1}L_\ell(\mu)P_s^\mathrm{base}(k, \mu)\ ,
\end{equation}
with $L_\ell$ being the $\ell$th order Legendre polynomial. We can then create nonlinear boost component emulators for each multipole $B_\ell(k) = P_\ell(k) \ / \ P^\mathrm{base}_\ell(k)$. 
The effect of RSD on small scales will already be included in the simulated $P_\ell(k)$ such that the neglected small scale RSD in the base model is captured by the prediction of the boost $B_\ell(k)$
\begin{equation}
    B_\ell(k) = B_{\ell,\mathrm{NL}}(k) + c_\ell(k)\ ,
\end{equation}
where $B_{\ell,\mathrm{NL}}(k)$ would be the scale dependant nonlinear boost, and $c_\ell(k)$ includes corrections to the base model that accommodate the neglected effects that predominately impact small scales. It should be noted that in order to obtain the best possible prediction accuracy we would want to keep $B_\ell(k)$, and thus $c_\ell(k)$, as small as possible. Exactly what small scale effects to include in the base model will depend on the given application of \texttt{matryoshka} and is beyond the scope of this work.

\subsection{\texttt{matryoshka} Python Package}
\label{subsec:python_package}
Alongside this paper we also publish the \texttt{matryoshka} Python package\footnote{\url{https://matryoshka-emu.readthedocs.io/en/latest/}}. This package includes all the weights for the NNs discussed in this paper, allowing them to be used without any requirement of re-training. The Python package has been developed such that the component emulators can be used in isolation. For example the transfer function component emulator can be loaded with \texttt{matryoshka.emulator.Transfer()}, and predictions can then be made with the \texttt{.emu\_predict()} method. The transfer function emulator makes predictions in $\sim 0.0004 \ \mathrm{s}$, and a nonlinear galaxy power spectrum prediction can be made in $\sim 0.1 \ \mathrm{s}$ (this is $\sim 3 \times$ faster than a transfer function prediction that can be made using \texttt{CLASS} with the accuracy settings implemented in \texttt{nbodykit}). It should be noted that although we have focused on training \texttt{matryoshka} to be used alongside the Aemulus simulations, it is simple to re-train any of the base model component emulators based on different parameters spaces. Functions to generate training samples and re-train the component emulators will be provided in the \texttt{matryoshka} Python package.

Many of the halo model functions in \texttt{matryoshka} are modified versions of those from the Python package \texttt{halomod} \citep{murray_thehalomod_2020}. In most cases these functions have been modified to allow \texttt{matryoshka} to make batch predictions more easily.

%% file: sections/conclusions.tex
We have introduced \texttt{matryoshka}, a suite of NN based emulators and Python package, that aims to produce fast and accurate predictions for the nonlinear galaxy power spectrum. The suite of emulators consists of a nonlinear boost component emulator along with four base model component emulators, allowing us to rapidly produce linear predictions of the galaxy power spectrum to be combined with predictions of a nonlinear boost.

In this paper we have demonstrated how the base model component emulators can be trained to be used along side a suite of numerical simulations, those being the Aemulus simulations. When trained on 6400 samples coming from the same 7D $w$CDM parameter space of the Aemulus simulations all base model component emulators have a MAPE < 0.02\%\footnote{This scalar version of the MAPE is calculated by taking the average prediction error of the test set, and then averaging this across all scales predicted by the component emulators} at 68\% CI (for scale dependent error of the base model component emulators see figure \ref{fig:linear_accu_comp}). The component emulators are capable of producing very fast predictions. The transfer function component emulator is capable of producing predictions in  $\sim0.0004 \ \mathrm{s}$ (the prediction time of the other component emulators is similar). This is $\sim 200\times$ faster than \texttt{CLASS} with the accuracy setting implemented via \texttt{nbodykit}. This speed up is more modest than that reported by other NN based linear emulators such as \citet{arico_accelerating_2021} or \citet{mancini_itcosmopower_2021}. Although it is difficult to quantitatively compare the prediction speed of these emulators with \texttt{matryoshka} without using consistent hardware and accuracy settings for Boltzmann codes that the predictions are being compared to. It is however likely that these other linear emulators produce faster predictions because the use of ensembling to reduce generalisation error in \texttt{matryoshka} does increase prediction time. Alternatives to ensembling will be explored in future works. The pre-trained transfer function emulator (and the other component emulators) is available in the \texttt{matryoshka} repository \url{https://github.com/JDonaldM/Matryoshka}.

Using \texttt{matryoshka} we investigated the potential gain in constraining power by including small nonlinear scales in a FS analysis of the galaxy power spectrum. To carry out this investigation we trained the nonlinear boost component emulator with nonlinear boosts calculated using HALOFIT. We approximate the scenario of using the Aemulus suite by only generating data for the 40 Aemulus training cosmologies and introducing simulation like noise into our training set. This nonlinear boost component emulator returns predictions with a MAPE <0.25\% at 68\% CI when evaluated with a test set produced for the Aemulus test samples (for scale dependent error from the nonlinear boost component emulator see figure \ref{fig:boost_accu}). Using \texttt{matryoshka} we conducted a series of FS analyses with different analysis setups on a mock power spectrum with underlying cosmology and HOD coming from \citet{planck_collaboration_planck_2020} and \citet{white_clustering_2011} respectively. From these analyses we estimate a $\sim 1.8 \times$ improvement in the constraint on $\sigma_8$ by increasing $k_\mathrm{max}$ from $0.25 \ h \ \mathrm{Mpc}^{-1}$ to $0.85 \ h \ \mathrm{Mpc}^{-1}$. This increases to the $\sim 4.8 \times$ improvement when the underlying HOD is known. These results highlight the potential gain in understanding we can achieve by using emulators in cosmological analyses, as well as motivating further studies of galaxy formation physics  that will inform us about appropriate HOD parameters.

%% file: sections/acknowledgement.tex
KK is supported the UK STFC grant ST/S000550/1. He was also supported by the European Research Council under the European Union's Horizon 2020 programme (grant agreement No.646702 ``CosTesGrav"). FB has received funding from the European Research Council (ERC) under the European Union's Horizon 2020 research and innovation programme (grant agreement 853291, ``FutureLSS"). FB is a Royal Society University Research Fellow. JDM was supported by a STFC studentship.

%% file: sections/data_av.tex
All training, validation, and test data for the \texttt{matryoshka} base model component emulators is available in the \texttt{matryoshka} repository (validation data was not produced for the nonlinear boost component emulator as such it is only available for the base model components). All weights for the NNs are also available in the repository. The Aemulus training and test samples can be found at \url{https://github.com/zxzhai/emulator}.

%% file: sections/convergence.tex
For this work we run our chains in two stages; a burn-in stage and a dense sampling stage. The burn-in stage uses the minimum number of walkers required by \texttt{zeus} ($2\times$ the number of dimensions). A small number of walkers is used here to allow for the walkers to move larger distances during each step of the chain, which results in a shorter burn in period. Every 200 steps we estimate the integrated autocorrelation time (IAT). We consider the burn-in stage complete once the total number of steps exceeds $5\times$ the IAT and our estimate of the IAT has changed by < 1\% compared to the previous estimate. For the dense sampling stage we increase the number of walkers by a factor of 6, so for our fiducial analysis this increases the number of walkers from 16 to 108. We run this dense sampling stage for $10\times$ the IAT estimate from the burn-in stage. The burn-in stage is discarded and the dense sampling stage is used for inference. Running the chains in this way allows us to exploit the rapid predictions from \texttt{matryoshka} whilst avoiding the long burn-in period that comes with using a large number of walkers with ensemble sampling.

%% file: sections/requirements.tex
\begin{figure}
	\includegraphics[width=\columnwidth]{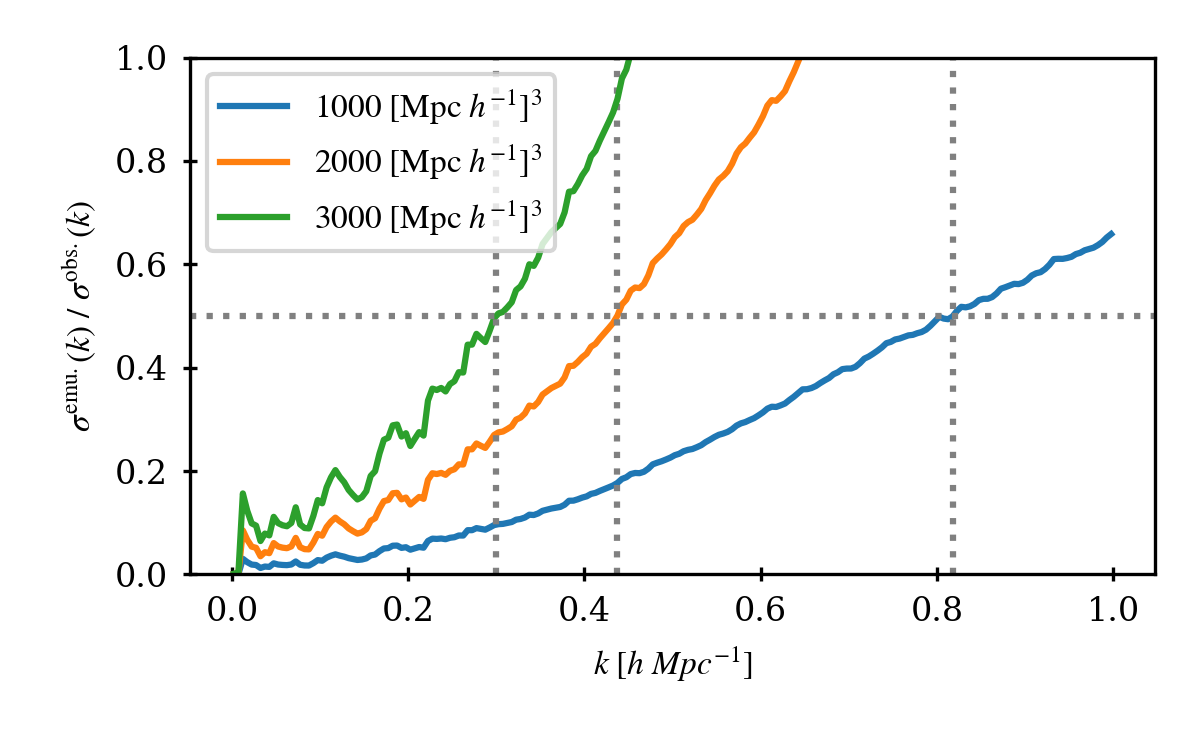}
    \caption{Ratio of the emulator error and statistical error for a CMASS-like sample with various sample volumes. The statistical errors are calculated using equation \ref{eq:gauss_cov}. The horizontal and vertical dotted grey lines indicate the scale at which the ratio of error equals 0.5.}
    \label{fig:emu_vs_stat_err}
\end{figure}

When developing an emulator we want to be able to produce predictions with errors that are smaller than the statistical errors of our observation, allowing us to return unbiased constraints on the parameters of interest within the statistical error. With this in mind, the statistical error of a given observation defines the accuracy requirement of the emulator predictions. For this work we have demonstrated training the nonlinear boost component emulator with the goal of conducting a series of FS analyses of a CMASS-like power spectrum with a sample volume of $1 \ (\mathrm{Gpc} \ h^{-1})^3$. Figure \ref{fig:emu_vs_stat_err} shows how the achieved level of prediction accuracy from \texttt{matryoshka} in this work compares to the statistical error from a CMASS-like sample with a volume of $1 \ (\mathrm{Gpc} \ h^{-1})^3$ (solid blue line). We can see that for $0.01 \ h \ \mathrm{Mpc}^{-1} \lesssim k \lesssim 1 \ h \ \mathrm{Mpc}^{-1}$ the statistical error is larger than the prediction error, however if we double the volume of the sample (shown by the solid orange line) this is no longer the case. In section \ref{sec:fullshape} we show that the achieved level of prediction accuracy is sufficient to return unbiased cosmological parameters when analysing a CMASS-like power spectrum up to $k=0.85 \ h \ \mathrm{Mpc}^{-1}$ with a sample volume of $1 \ (\mathrm{Gpc} \ h^{-1})^3$. This unlikely to be the case for a sample volume of $2 \ (\mathrm{Gpc} \ h^{-1})^3$ as the emulator error is going to be larger than the statistical error on those scales as seen in figure \ref{fig:emu_vs_stat_err}.

%% file: main.bbl
\begin{thebibliography}{}
\makeatletter
\relax
\def\mn@urlcharsother{\let\do\@makeother \do\$\do\&\do\#\do\^\do\_\do\%\do\~}
\def\mn@doi{\begingroup\mn@urlcharsother \@ifnextchar [ {\mn@doi@}
  {\mn@doi@[]}}
\def\mn@doi@[#1]#2{\def\@tempa{#1}\ifx\@tempa\@empty \href
  {http://dx.doi.org/#2} {doi:#2}\else \href {http://dx.doi.org/#2} {#1}\fi
  \endgroup}
\def\mn@eprint#1#2{\mn@eprint@#1:#2::\@nil}
\def\mn@eprint@arXiv#1{\href {http://arxiv.org/abs/#1} {{\tt arXiv:#1}}}
\def\mn@eprint@dblp#1{\href {http://dblp.uni-trier.de/rec/bibtex/#1.xml}
  {dblp:#1}}
\def\mn@eprint@#1:#2:#3:#4\@nil{\def\@tempa {#1}\def\@tempb {#2}\def\@tempc
  {#3}\ifx \@tempc \@empty \let \@tempc \@tempb \let \@tempb \@tempa \fi \ifx
  \@tempb \@empty \def\@tempb {arXiv}\fi \@ifundefined
  {mn@eprint@\@tempb}{\@tempb:\@tempc}{\expandafter \expandafter \csname
  mn@eprint@\@tempb\endcsname \expandafter{\@tempc}}}

\bibitem[\protect\citeauthoryear{Abadi et~al.,}{Abadi
  et~al.}{2016}]{abadi_tensorflow_2016}
Abadi M.,  et~al., 2016, {arXiv}:1603.04467

\bibitem[\protect\citeauthoryear{Agarwal, Abdalla, Feldman, Lahav  \&
  Thomas}{Agarwal et~al.}{2012}]{agarwal_pkann_2012}
Agarwal S.,  Abdalla F.~B.,  Feldman H.~A.,  Lahav O.,   Thomas S.~A.,  2012,
  \mn@doi [\mnras] {10.1111/j.1365-2966.2012.21326.x}, 424, 1409

\bibitem[\protect\citeauthoryear{Agarwal, Abdalla, Feldman, Lahav  \&
  Thomas}{Agarwal et~al.}{2014}]{agarwal_pkann_2014}
Agarwal S.,  Abdalla F.~B.,  Feldman H.~A.,  Lahav O.,   Thomas S.~A.,  2014,
  \mn@doi [\mnras] {10.1093/mnras/stu090}, 439, 2102

\bibitem[\protect\citeauthoryear{Aghamousa et~al.,}{Aghamousa
  et~al.}{2016}]{desi_collaboration_desi_2016}
Aghamousa A.,  et~al., 2016, {arXiv}:1611.00036

\bibitem[\protect\citeauthoryear{Aghanim et~al.,}{Aghanim
  et~al.}{2020}]{planck_collaboration_planck_2020}
Aghanim N.,  et~al., 2020, \mn@doi [{A\&A}] {10.1051/0004-6361/201833910}, 641

\bibitem[\protect\citeauthoryear{Alam et~al.,}{Alam
  et~al.}{2017}]{alam_clustering_2017}
Alam S.,  et~al., 2017, \mn@doi [\mnras] {10.1093/mnras/stx721}, 470, 2617

\bibitem[\protect\citeauthoryear{Alam et~al.,}{Alam
  et~al.}{2021}]{eboss_collaboration_completed_2021}
Alam S.,  et~al., 2021, \mn@doi [\prd] {10.1103/PhysRevD.103.083533}, 103,
  083533

\bibitem[\protect\citeauthoryear{Alsing et~al.,}{Alsing
  et~al.}{2020}]{alsing_speculator_2020}
Alsing J.,  et~al., 2020, \mn@doi [{ApJS}] {10.3847/1538-4365/ab917f}, 249, 5

\bibitem[\protect\citeauthoryear{Angulo \& Pontzen}{Angulo \&
  Pontzen}{2016}]{angulo_cosmological_2016}
Angulo R.~E.,  Pontzen A.,  2016, \mn@doi [\mnras] {10.1093/mnrasl/slw098},
  462, L1

\bibitem[\protect\citeauthoryear{Angulo, Zennaro, Contreras, Aricò,
  Pellejero-Ibañez  \& Stücker}{Angulo et~al.}{2020}]{angulo_bacco_2020}
Angulo R.~E.,  Zennaro M.,  Contreras S.,  Aricò G.,  Pellejero-Ibañez M.,
  Stücker J.,  2020, {arXiv}:2004.06245

\bibitem[\protect\citeauthoryear{Aricò, Angulo, Hernández-Monteagudo,
  Contreras, Zennaro, Pellejero-Ibañez  \& Rosas-Guevara}{Aricò
  et~al.}{2020}]{arico_modelling_2020}
Aricò G.,  Angulo R.~E.,  Hernández-Monteagudo C.,  Contreras S.,  Zennaro
  M.,  Pellejero-Ibañez M.,   Rosas-Guevara Y.,  2020, \mn@doi [\mnras]
  {10.1093/mnras/staa1478}, 495, 4800

\bibitem[\protect\citeauthoryear{Aricò, Angulo  \& Zennaro}{Aricò
  et~al.}{2021}]{arico_accelerating_2021}
Aricò G.,  Angulo R.~E.,   Zennaro M.,  2021, {arXiv}:2104.14568

\bibitem[\protect\citeauthoryear{Beutler et~al.,}{Beutler
  et~al.}{2013}]{beutler_6df_2013}
Beutler F.,  et~al., 2013, \mn@doi [\mnras] {10.1093/mnras/sts637}, 429, 3604

\bibitem[\protect\citeauthoryear{Bird, Rogers, Peiris, Verde, Font-Ribera  \&
  Pontzen}{Bird et~al.}{2019}]{bird_emulator_2019}
Bird S.,  Rogers K.~K.,  Peiris H.~V.,  Verde L.,  Font-Ribera A.,   Pontzen
  A.,  2019, \mn@doi [\jcap] {10.1088/1475-7516/2019/02/050}, 2019, 050

\bibitem[\protect\citeauthoryear{Chapman et~al.,}{Chapman
  et~al.}{2021}]{chapman_completed_2021}
Chapman M.~J.,  et~al., 2021, {arXiv}:2106.14961

\bibitem[\protect\citeauthoryear{Chuang et~al.,}{Chuang
  et~al.}{2019}]{chuang_unit_2019}
Chuang C.-H.,  et~al., 2019, \mn@doi [\mnras] {10.1093/mnras/stz1233}, 487, 48

\bibitem[\protect\citeauthoryear{Cole et~al.,}{Cole
  et~al.}{2005}]{cole_2df_2005}
Cole S.,  et~al., 2005, \mn@doi [\mnras] {10.1111/j.1365-2966.2005.09318.x},
  362, 505

\bibitem[\protect\citeauthoryear{Cooray \& Sheth}{Cooray \&
  Sheth}{2002}]{cooray_halo_2002}
Cooray A.,  Sheth R.,  2002, \mn@doi [\physrep]
  {10.1016/S0370-1573(02)00276-4}, 372, 1

\bibitem[\protect\citeauthoryear{Dawson et~al.,}{Dawson
  et~al.}{2012}]{dawson_baryon_2012}
Dawson K.~S.,  et~al., 2012, \mn@doi [\aj] {10.1088/0004-6256/145/1/10}, 145,
  10

\bibitem[\protect\citeauthoryear{{DeRose} et~al.,}{{DeRose}
  et~al.}{2019}]{derose_aemulus_2019}
{DeRose} J.,  et~al., 2019, \mn@doi [\apj] {10.3847/1538-4357/ab1085}, 875, 69

\bibitem[\protect\citeauthoryear{Debackere, Schaye  \& Hoekstra}{Debackere
  et~al.}{2020}]{debackere_impact_2020}
Debackere S. N.~B.,  Schaye J.,   Hoekstra H.,  2020, \mn@doi [\mnras]
  {10.1093/mnras/stz3446}, 492, 2285

\bibitem[\protect\citeauthoryear{Dolag, Borgani, Schindler, Diaferio  \&
  Bykov}{Dolag et~al.}{2008}]{dolag_simulation_2008}
Dolag K.,  Borgani S.,  Schindler S.,  Diaferio A.,   Bykov A.~M.,  2008,
  \mn@doi [\ssr] {10.1007/s11214-008-9316-5}, 134, 229

\bibitem[\protect\citeauthoryear{Duffy, Schaye, Kay  \& Vecchia}{Duffy
  et~al.}{2008}]{duffy_dark_2008}
Duffy A.~R.,  Schaye J.,  Kay S.~T.,   Vecchia C.~D.,  2008, \mn@doi [\mnras]
  {10.1111/j.1745-3933.2008.00537.x}, 390, L64

\bibitem[\protect\citeauthoryear{Eisenstein \& Hu}{Eisenstein \&
  Hu}{1998}]{eisenstein_baryonic_1998}
Eisenstein D.~J.,  Hu W.,  1998, \mn@doi [\apj] {10.1086/305424}, 496, 605

\bibitem[\protect\citeauthoryear{Foreman, Perrier  \& Senatore}{Foreman
  et~al.}{2016}]{foreman_precision_2016}
Foreman S.,  Perrier H.,   Senatore L.,  2016, \mn@doi [\jcap]
  {10.1088/1475-7516/2016/05/027}, 2016, 027

\bibitem[\protect\citeauthoryear{Garrison, Eisenstein, Ferrer, Tinker, Pinto
  \& Weinberg}{Garrison et~al.}{2018}]{garrison_abacus_2018}
Garrison L.~H.,  Eisenstein D.~J.,  Ferrer D.,  Tinker J.~L.,  Pinto P.~A.,
  Weinberg D.~H.,  2018, \mn@doi [{ApJS}] {10.3847/1538-4365/aabfd3}, 236, 43

\bibitem[\protect\citeauthoryear{Giblin, Cataneo, Moews  \& Heymans}{Giblin
  et~al.}{2019}]{giblin_road_2019}
Giblin B.,  Cataneo M.,  Moews B.,   Heymans C.,  2019, \mn@doi [\mnras]
  {10.1093/mnras/stz2659}, 490, 4826

\bibitem[\protect\citeauthoryear{Habib, Heitmann, Higdon, Nakhleh  \&
  Williams}{Habib et~al.}{2007}]{habib_cosmic_2007}
Habib S.,  Heitmann K.,  Higdon D.,  Nakhleh C.,   Williams B.,  2007, \mn@doi
  [\prd] {10.1103/PhysRevD.76.083503}, 76, 083503

\bibitem[\protect\citeauthoryear{Hamilton}{Hamilton}{2000}]{hamilton_uncorrelated_2000}
Hamilton A. J.~S.,  2000, \mn@doi [\mnras] {10.1046/j.1365-8711.2000.03071.x},
  312, 257

\bibitem[\protect\citeauthoryear{Hand, Feng, Beutler, Li, Modi, Seljak  \&
  Slepian}{Hand et~al.}{2018}]{Hand_2018}
Hand N.,  Feng Y.,  Beutler F.,  Li Y.,  Modi C.,  Seljak U.,   Slepian Z.,
  2018, \mn@doi [\apj] {10.3847/1538-3881/aadae0}, 156, 160

\bibitem[\protect\citeauthoryear{Heitmann, Higdon, White, Habib, Williams,
  Lawrence  \& Wagner}{Heitmann et~al.}{2009}]{heitmann_coyote_2009}
Heitmann K.,  Higdon D.,  White M.,  Habib S.,  Williams B.~J.,  Lawrence E.,
  Wagner C.,  2009, \mn@doi [\apj] {10.1088/0004-637X/705/1/156}, 705, 156

\bibitem[\protect\citeauthoryear{Ivanov, Simonović  \& Zaldarriaga}{Ivanov
  et~al.}{2020}]{ivanov_cosmological_2020}
Ivanov M.~M.,  Simonović M.,   Zaldarriaga M.,  2020, \mn@doi [\jcap]
  {10.1088/1475-7516/2020/05/042}, 2020, 042

\bibitem[\protect\citeauthoryear{Karamanis, Beutler  \& Peacock}{Karamanis
  et~al.}{2021}]{karamanis_zeus_2021}
Karamanis M.,  Beutler F.,   Peacock J.~A.,  2021, {arXiv}:2105.03468

\bibitem[\protect\citeauthoryear{Kingma \& Ba}{Kingma \&
  Ba}{2017}]{kingma_adam:_2017}
Kingma D.~P.,  Ba J.,  2017, {arXiv}:1412.6980

\bibitem[\protect\citeauthoryear{Klypin, Prada  \& Byun}{Klypin
  et~al.}{2020}]{klypin_suppressing_2020}
Klypin A.,  Prada F.,   Byun J.,  2020, \mn@doi [\mnras]
  {10.1093/mnras/staa734}, 496, 3862

\bibitem[\protect\citeauthoryear{Knabenhans et~al.,}{Knabenhans
  et~al.}{2019}]{euclid_collaboration_euclid_2019}
Knabenhans M.,  et~al., 2019, \mn@doi [MNRAS] {10.1093/mnras/stz197}, 484, 5509

\bibitem[\protect\citeauthoryear{Knabenhans et~al.,}{Knabenhans
  et~al.}{2020}]{euclid_collaboration_euclid_2020}
Knabenhans M.,  et~al., 2020, {arXiv}:2010.11288

\bibitem[\protect\citeauthoryear{Kobayashi, Nishimichi, Takada, Takahashi  \&
  Osato}{Kobayashi et~al.}{2020}]{kobayashi_accurate_2020}
Kobayashi Y.,  Nishimichi T.,  Takada M.,  Takahashi R.,   Osato K.,  2020,
  \mn@doi [\prd] {10.1103/PhysRevD.102.063504}, 102, 063504

\bibitem[\protect\citeauthoryear{Kuhlen, Vogelsberger  \& Angulo}{Kuhlen
  et~al.}{2012}]{kuhlen_numerical_2012}
Kuhlen M.,  Vogelsberger M.,   Angulo R.,  2012, \mn@doi [Physics of the Dark
  Universe] {10.1016/j.dark.2012.10.002}, 1, 50

\bibitem[\protect\citeauthoryear{Kwan, Heitmann, Habib, Padmanabhan, Finkel,
  Frontiere  \& Pope}{Kwan et~al.}{2015}]{kwan_cosmic_2015}
Kwan J.,  Heitmann K.,  Habib S.,  Padmanabhan N.,  Finkel H.,  Frontiere N.,
  Pope A.,  2015, \mn@doi [{ApJ}] {10.1088/0004-637X/810/1/35}, 810, 35

\bibitem[\protect\citeauthoryear{Lange, Hearin, Leauthaud, Bosch, Guo  \&
  {DeRose}}{Lange et~al.}{2021}]{lange_five-percent_2021}
Lange J.~U.,  Hearin A.~P.,  Leauthaud A.,  Bosch F. C. v.~d.,  Guo H.,
  {DeRose} J.,  2021, {arXiv}:2101.12261

\bibitem[\protect\citeauthoryear{Laureijs et~al.,}{Laureijs
  et~al.}{2011}]{laureijs_euclid_2011}
Laureijs R.,  et~al., 2011, {arXiv}:1110.3193

\bibitem[\protect\citeauthoryear{Lawrence et~al.,}{Lawrence
  et~al.}{2017}]{lawrence_mira-titan_2017}
Lawrence E.,  et~al., 2017, \mn@doi [\apj] {10.3847/1538-4357/aa86a9}, 847, 50

\bibitem[\protect\citeauthoryear{Lesgourgues}{Lesgourgues}{2011}]{lesgourgues_cosmic_2011}
Lesgourgues J.,  2011, {arXiv}:1104.2932

\bibitem[\protect\citeauthoryear{Levi et~al.,}{Levi
  et~al.}{2013}]{levi_desi_2013}
Levi M.,  et~al., 2013, {arXiv}:1308.0847

\bibitem[\protect\citeauthoryear{Lewis, Challinor  \& Lasenby}{Lewis
  et~al.}{2000}]{lewis_efficient_2000}
Lewis A.,  Challinor A.,   Lasenby A.,  2000, \mn@doi [\apj] {10.1086/309179},
  538, 473

\bibitem[\protect\citeauthoryear{Mancini, Piras, Alsing, Joachimi  \&
  Hobson}{Mancini et~al.}{2021}]{mancini_itcosmopower_2021}
Mancini A.~S.,  Piras D.,  Alsing J.,  Joachimi B.,   Hobson M.~P.,  2021,
  {arXiv}:2106.03846

\bibitem[\protect\citeauthoryear{Miyatake et~al.,}{Miyatake
  et~al.}{2020}]{miyatake_cosmological_2020}
Miyatake H.,  et~al., 2020, {arXiv}:2101.00113

\bibitem[\protect\citeauthoryear{Mootoovaloo, Jaffe, Heavens  \&
  Leclercq}{Mootoovaloo et~al.}{2021}]{mootoovaloo_kernel-based_2021}
Mootoovaloo A.,  Jaffe A.~H.,  Heavens A.~F.,   Leclercq F.,  2021,
  {arXiv}:2105.02256

\bibitem[\protect\citeauthoryear{Murray, Power  \& Robotham}{Murray
  et~al.}{2013}]{murray_hmfcalc_2013}
Murray S.,  Power C.,   Robotham A.,  2013, {arXiv}:1306.6721

\bibitem[\protect\citeauthoryear{Murray, Diemer  \& Chen}{Murray
  et~al.}{2020}]{murray_thehalomod_2020}
Murray S.~G.,  Diemer B.,   Chen Z.,  2020, {arXiv}:2009.14066

\bibitem[\protect\citeauthoryear{Navarro, Frenk  \& White}{Navarro
  et~al.}{1996}]{navarro_structure_1996}
Navarro J.~F.,  Frenk C.~S.,   White S. D.~M.,  1996, \mn@doi [\apj]
  {10.1086/177173}, 462, 563

\bibitem[\protect\citeauthoryear{Nishimichi et~al.,}{Nishimichi
  et~al.}{2019}]{nishimichi_dark_2019}
Nishimichi T.,  et~al., 2019, \mn@doi [\apj] {10.3847/1538-4357/ab3719}, 884,
  29

\bibitem[\protect\citeauthoryear{Parejko et~al.,}{Parejko
  et~al.}{2013}]{parejko_clustering_2013}
Parejko J.~K.,  et~al., 2013, \mn@doi [MNRAS] {10.1093/mnras/sts314}, 429, 98

\bibitem[\protect\citeauthoryear{Pedersen, Font-Ribera, Rogers, {McDonald},
  Peiris, Pontzen  \& Slosar}{Pedersen et~al.}{2021}]{pedersen_emulator_2021}
Pedersen C.,  Font-Ribera A.,  Rogers K.~K.,  {McDonald} P.,  Peiris H.~V.,
  Pontzen A.,   Slosar A.,  2021, \mn@doi [\jcap]
  {10.1088/1475-7516/2021/05/033}, 2021, 033

\bibitem[\protect\citeauthoryear{Percival, Cole, Eisenstein, Nichol, Peacock,
  Pope  \& Szalay}{Percival et~al.}{2007}]{percival_measuring_2007}
Percival W.~J.,  Cole S.,  Eisenstein D.~J.,  Nichol R.~C.,  Peacock J.~A.,
  Pope A.~C.,   Szalay A.~S.,  2007, \mn@doi [\mnras]
  {10.1111/j.1365-2966.2007.12268.x}, 381, 1053

\bibitem[\protect\citeauthoryear{Philcox, Ivanov, Simonović  \&
  Zaldarriaga}{Philcox et~al.}{2020}]{philcox_combining_2020}
Philcox O.~H.,  Ivanov M.~M.,  Simonović M.,   Zaldarriaga M.,  2020, \mn@doi
  [\jcap] {10.1088/1475-7516/2020/05/032}, 2020, 032

\bibitem[\protect\citeauthoryear{Schneider et~al.,}{Schneider
  et~al.}{2016}]{schneider_matter_2016}
Schneider A.,  et~al., 2016, \mn@doi [\jcap] {10.1088/1475-7516/2016/04/047},
  2016, 047

\bibitem[\protect\citeauthoryear{Schneider, Teyssier, Stadel, Chisari, Brun,
  Amara  \& Refregier}{Schneider et~al.}{2019}]{schneider_quantifying_2019}
Schneider A.,  Teyssier R.,  Stadel J.,  Chisari N.~E.,  Brun A. M.~L.,  Amara
  A.,   Refregier A.,  2019, \mn@doi [\jcap] {10.1088/1475-7516/2019/03/020},
  2019, 020

\bibitem[\protect\citeauthoryear{Senatore}{Senatore}{2015}]{senatore_bias_2015}
Senatore L.,  2015, \mn@doi [\jcap] {10.1088/1475-7516/2015/11/007}, 2015, 007

\bibitem[\protect\citeauthoryear{Senatore \& Zaldarriaga}{Senatore \&
  Zaldarriaga}{2015}]{senatore_ir-resummed_2015}
Senatore L.,  Zaldarriaga M.,  2015, \mn@doi [\jcap]
  {10.1088/1475-7516/2015/02/013}, 2015, 013

\bibitem[\protect\citeauthoryear{Simonović, Baldauf, Zaldarriaga, Carrasco  \&
  Kollmeier}{Simonović et~al.}{2018}]{simonovic_cosmological_2018}
Simonović M.,  Baldauf T.,  Zaldarriaga M.,  Carrasco J.~J.,   Kollmeier
  J.~A.,  2018, \mn@doi [\jcap] {10.1088/1475-7516/2018/04/030}, 2018, 030

\bibitem[\protect\citeauthoryear{Takahashi, Sato, Nishimichi, Taruya  \&
  Oguri}{Takahashi et~al.}{2012}]{takahashi_revising_2012}
Takahashi R.,  Sato M.,  Nishimichi T.,  Taruya A.,   Oguri M.,  2012, \mn@doi
  [\apj] {10.1088/0004-637X/761/2/152}, 761, 152

\bibitem[\protect\citeauthoryear{TensorFlow}{TensorFlow}{2021}]{tensorflow_developers_tensorflow_2021}
TensorFlow 2021, {TensorFlow}, \mn@doi{10.5281/ZENODO.4724125}, \url
  {https://zenodo.org/record/4724125}

\bibitem[\protect\citeauthoryear{Tinker, Kravtsov, Klypin, Abazajian, Warren,
  Yepes, Gottlober  \& Holz}{Tinker et~al.}{2008}]{tinker_toward_2008}
Tinker J.~L.,  Kravtsov A.~V.,  Klypin A.,  Abazajian K.,  Warren M.~S.,  Yepes
  G.,  Gottlober S.,   Holz D.~E.,  2008, \mn@doi [\apj] {10.1086/591439}, 688,
  709

\bibitem[\protect\citeauthoryear{Tinker, Robertson, Kravtsov, Klypin, Warren,
  Yepes  \& Gottlober}{Tinker et~al.}{2010}]{tinker_large_2010}
Tinker J.~L.,  Robertson B.~E.,  Kravtsov A.~V.,  Klypin A.,  Warren M.~S.,
  Yepes G.,   Gottlober S.,  2010, \mn@doi [\apj]
  {10.1088/0004-637X/724/2/878}, 724, 878

\bibitem[\protect\citeauthoryear{Villaescusa-Navarro
  et~al.,}{Villaescusa-Navarro et~al.}{2020}]{villaescusa-navarro_quijote_2020}
Villaescusa-Navarro F.,  et~al., 2020, \mn@doi [{ApJS}]
  {10.3847/1538-4365/ab9d82}, 250, 2

\bibitem[\protect\citeauthoryear{Vogelsberger, Marinacci, Torrey  \&
  Puchwein}{Vogelsberger et~al.}{2019}]{vogelsberger_cosmological_2019}
Vogelsberger M.,  Marinacci F.,  Torrey P.,   Puchwein E.,  2019,
  {arXiv}:1909.07976

\bibitem[\protect\citeauthoryear{White et~al.,}{White
  et~al.}{2011}]{white_clustering_2011}
White M.,  et~al., 2011, \mn@doi [{ApJ}] {10.1088/0004-637X/728/2/126}, 728,
  126

\bibitem[\protect\citeauthoryear{Zhai et~al.,}{Zhai
  et~al.}{2017}]{zhai_clustering_2017}
Zhai Z.,  et~al., 2017, \mn@doi [\apj] {10.3847/1538-4357/aa8eee}, 848, 76

\bibitem[\protect\citeauthoryear{Zhai et~al.,}{Zhai
  et~al.}{2019}]{zhai_aemulus_2019}
Zhai Z.,  et~al., 2019, \mn@doi [{ApJ}] {10.3847/1538-4357/ab0d7b}, 874, 95

\bibitem[\protect\citeauthoryear{Zheng et~al.,}{Zheng
  et~al.}{2005}]{zheng_theoretical_2005}
Zheng Z.,  et~al., 2005, \mn@doi [\apj] {10.1086/466510}, 633, 791

\bibitem[\protect\citeauthoryear{Zheng, Coil  \& Zehavi}{Zheng
  et~al.}{2007}]{zheng_galaxy_2007}
Zheng Z.,  Coil A.~L.,   Zehavi I.,  2007, \mn@doi [\apj] {10.1086/521074},
  667, 760

\bibitem[\protect\citeauthoryear{Zheng, Zehavi, Eisenstein, Weinberg  \&
  Jing}{Zheng et~al.}{2009}]{zheng_halo_2009}
Zheng Z.,  Zehavi I.,  Eisenstein D.~J.,  Weinberg D.~H.,   Jing Y.~P.,  2009,
  \mn@doi [{ApJ}] {10.1088/0004-637X/707/1/554}, 707, 554

\bibitem[\protect\citeauthoryear{Zhou et~al.,}{Zhou
  et~al.}{2020}]{zhou_clustering_2020}
Zhou R.,  et~al., 2020, {arXiv}:2001.06018

\makeatother
\end{thebibliography}
